\documentclass[letterpaper,prc,twocolumn,showpacs,floatfix,nofootinbib,preprintnumbers,superscriptaddress,amsmath,amssymb]{revtex4-1}

\usepackage{amsmath}           
\usepackage{amsfonts}
\usepackage{graphicx}          
\usepackage{dcolumn}           
\usepackage{bm}
\usepackage{color}
\usepackage[english]{babel}
\usepackage{graphicx}

\usepackage{physicsShortCut}
\usepackage{tdgcmNotations}

\newcommand{\gras}[1]{\boldsymbol{#1}}

\begin{document}

\title{Fission fragment charge and mass distributions in $^{239}$Pu(n,f) in the adiabatic nuclear energy density functional theory}

\author{D.~Regnier}
\email[]{david.regnier@cea.fr}
\affiliation{CEA,DAM,DIF, 91297 Arpajon, France}

\author{N.~Dubray}
\email[]{noel.dubray@cea.fr}
\affiliation{CEA,DAM,DIF, 91297 Arpajon, France}

\author{N.~Schunck}
\email[]{schunck1@llnl.gov}
\affiliation{Nuclear and Chemical Science Division, LLNL, Livermore, CA 94551, USA}

\author{M.~Verri\`ere}
\affiliation{CEA,DAM,DIF, 91297 Arpajon, France}

\date{\today}

\begin{abstract}
\begin{description}
\item[Background]
Accurate knowledge of fission fragment yields is an essential ingredient of numerous applications ranging from the formation of elements in the r-process to fuel cycle optimization for nuclear energy. The need for a predictive theory applicable where no data is available, together with the variety of potential applications, is an incentive to develop a fully microscopic approach to fission dynamics.
\item[Purpose]
In this work, we calculate the pre-neutron emission charge and mass distributions of the fission fragments formed in the neutron-induced fission of $^{239}$Pu using a microscopic method based on nuclear density functional theory (DFT).
\item[Methods]
Our theoretical framework is the nuclear energy density functional (EDF) method, where large amplitude collective motion is treated adiabatically using the time dependent generator coordinate method (TDGCM) under the Gaussian overlap approximation (GOA). In practice, the TDGCM is implemented in two steps. First, a series of constrained EDF calculations map the configuration and potential energy landscape of the fissioning system for a small set of collective variables (in this work, the axial quadrupole and octupole moments of the nucleus). Then, nuclear dynamics is modeled by propagating a collective wave packet on the potential energy surface. Fission fragment distributions are extracted from the flux of the collective wave packet through the scission line.
\item[Results]
We find that the main characteristics of the fission charge and mass distributions can be well reproduced by existing energy functionals even in two-dimensional collective spaces. Theory and experiment agree typically within 2 mass units for the position of the asymmetric peak. As expected, calculations are sensitive to the structure of the initial state and the prescription for the collective inertia. We emphasize that results are also sensitive to the continuity of the collective landscape near scission.
\item[Conclusions]
Our analysis confirms that the adiabatic approximation provides an effective scheme to compute fission fragment yields. It also suggests that, at least in the framework of nuclear DFT, three-dimensional collective spaces may be a prerequisite to reach 10\% accuracy in predicting pre-neutron emission fission fragment yields.
\end{description}
\end{abstract}


\maketitle


\section{\label{sec:intro} Introduction}

Nuclear fission is the mechanism under which a (usually heavy) atomic nucleus may split into two or more fragments, thereby releasing a large amount of energy \cite{krappe2012}. In addition to important applications in energy production or national security, fission also plays a crucial role in determining the stability of superheavy elements \cite{baran2015}. As a consequence, it is one of the primary mechanisms that terminate nucleosynthesis, and there have been speculations that fission fragments could be re-absorbed in the r-process flow \cite{goriely2011,korobkin2012,just2015}. Fission is also an important mechanism used to produce short-lived exotic nuclides at experimental facilities with radioactive ion beams, and is thus a portal to studying the frontiers of nuclear stability.

Although the mechanism of nuclear fission has been known for over 80 years, reliably predicting its characteristics solely on the basis of our knowledge of nuclear forces and quantum many-body methods remains an elusive goal \cite{schunck2016}. In common with most of nuclear structure theory, one of the main hurdles to developing such a predictive theory of fission is our relatively poor knowledge of nuclear forces. In addition, nuclear fission is a prime example of a non-equilibrium, time-dependent, large amplitude collective motion where a many-body system of strongly-interacting fermions decays through the coupling to a number of open channels. In light of these theoretical difficulties, it is quite remarkable that current fission theory can reproduce the 30 orders of magnitude range of experimentally measured spontaneous fission half-lives~\cite{staszczak2013,rodrigues_microscopic_2014} and predict the main characteristics of the mass yields in neutron-induced fission of actinides~\cite{younes2012}.

Today, there is a relative consensus that nuclear density functional theory provides the most promising framework to describe fission microscopically \cite{schunck2016}. By reformulating the original nuclear many-body problem in terms of an energy density functional of the isoscalar particle density $\rho$, it offers tremendous conceptual and practical simplifications. In particular, the full many-body Schr\"odinger equation reduces to a Hartree-Fock-Bogoliubov-like equation which is easily implemented on modern computers. In addition, the concept of nuclear deformation, which is a particular realization of the generic spontaneous symmetry breaking so essential to the success of the EDF approach, allows to cast bridges between microscopic and phenomenological models of fission. Historically, these models, where the nucleus is modeled as a deformed liquid drop with several corrections of quantum origin, have been extraordinarily successful in applications, see, e.g., \cite{nadtochy2005,nadtochy2007,randrup2011,randrup2011-a,sadhukhan2011,moeller2012,randrup2013} for a selection of some recent applications.

Nuclear DFT can be applied to fission in two different flavors. In time-dependent density functional theory (TDDFT), the real-time evolution of the nuclear system is simulated from an initial condition up to scission \cite{nakatsukasa2012,bulgac2013}. This approach automatically includes one-body dissipation effects: as the nucleus changes its shape, single-particle excitations (or quasi-particle excitations when pairing correlations are taken into account) are included, which slows the collective motion. TDDFT is probably the most promising approach to describe the structure of the fission fragments and various recent works have shown extremely encouraging results \cite{simenel2014,goddard2015,scamps2015,bulgac2016}. However, realistic simulation of a single fission event including full symmetry breaking and full treatment of pairing correlations is at the limit of what current supercomputers can handle \cite{bulgac2016}. Since predicting fission fragment distributions in this framework would probably require sampling thousands of such events, this approach is currently inapplicable.

By contrast, the adiabatic approximation to fission relies on the observation that fission timescales are much longer than the timescales of single-particle excitations. Typical estimates of the time it takes for the nucleus to go from saddle to scission is in the range of $10^{-19} - 10^{-21}$ s (the more realistic the model, the longer the timescale) \cite{negele1978,berger1986,berger1991,bulgac2016}; time-scales associated with Fermi energies of -8 MeV are $\tau_{\mathrm{sp}} = \hbar/\lambda_{\mathrm{F}}c \approx 10^{-22}$ s. As a result, one may assume equilibrium at all times and decouple the large amplitude motion of the system as a whole from internal excitations. In practice, this can be very effectively achieved by introducing a small set of collective variables such as multipole moments, pairing fields, etc., and precalculating the potential energy surface of the nucleus as a function of these variables. This strategy has proven especially fruitful to compute spontaneous fission half-lives through simple multi-dimensional quantum tunneling \cite{warda2002,warda2012,staszczak2005,schindzielorz2009,staszczak2013,rodriguez-guzman2014,rodriguez-guzman2014-a,sadhukhan2014,baran2015}. As we will show in this work, it is also the ideal framework to compute charge and mass distributions.

In this paper, we will use such an adiabatic approach to large amplitude collective motion based on the time-dependent generator coordinate method (TDGCM) to compute the fission fragment charge and mass distributions (before any neutron emission) for the benchmark case of $^{240}$Pu. While the TDGCM was already introduced as early as 1991 \cite{berger1991}, there have been very few applications to the calculation of fission fragment distributions \cite{goutte2005,younes2012-a}. Here, we will perform a comprehensive study by examining in details the role in the collective dynamics of the initial state and of the collective inertia. We will also estimate systematic uncertainties by comparing predictions from two different functionals. The purpose of this work is to establish a complete benchmark that can be used to gauge future progress.

In section \ref{sec:theory}, we briefly describe the main features of the theoretical approach. We recall the general ansatz of the TDGCM, discuss the calculation of the potential energy surface and of the time-evolution of the system. In section \ref{sec:results}, we analyze the results by discussing the numerical convergence of the calculation, the sensitivity to the initial state, to the collective inertia, to the EDF, etc. We conclude with a brief summary and outlook in section \ref{sec:conclusion}.


\section{\label{sec:theory} Theoretical framework}

This section presents the methodology adopted to determine a collective Schr\"odinger-like equation that describes low-energy fission dynamics. We also explain how to extract the fission fragment distributions from the time evolution of the resulting collective wave packet.


\subsection{Overview of the method}

In this work, fission is described as a slow adiabatic process driven by a few collective degrees of freedom, or collective variables. The time-dependent extension to the generator coordinate method provides an appropriate formalism to describe such a large amplitude motion in nuclei \cite{reinhard1987,ring2000}. In this approach, we assume that the many-body state $| \Psi(t) \rangle$ of the fissioning system takes the generic form
\begin{equation}
\label{eq:gcmApprox}
|\Psi(t)\rangle= \int_{\qvec} f(\qvec, t) |\Phi_{\qvec} \rangle \, \text{d}\qvec.
\end{equation}
The set $\{ |\Phi_{\qvec}\rangle \}_\qvec$ is made of known many-body states parametrized by a vector of continuous variables $\qvec \equiv (q_{1}, \dots, q_{N})$. Each of these $q_{i}$ is a collective variable and must be chosen based on the physics of the problem. Inserting the form (\ref{eq:gcmApprox}) in the time-dependent many-body Schr\"odinger equation, which governs the evolution of the fissioning nuclei, yields an equation for the unknown weight function $f(\qvec, t)$, the Hill-Wheeler equation,
\begin{equation}
 \forall \,\qvec: \quad \int_{\qpvec} \langle \Phi_{\qvec} | \left[ \hat{H} - i\hbar \frac{\partial }{\partial t}\right]|  \Phi_{\qpvec} \rangle f(\qpvec,t)=0.
\end{equation}
In this work, we will assume that the potential part of the nuclear Hamiltonian $\hat{H}$ is approximated by an effective two-body interaction of the Skyrme or Gogny type. Numerically solving the time-dependent Hill-Wheeler equation for fissioning systems would require a tremendous amount of computational resources and has not been attempted yet. Instead, a widespread approach consists in injecting an additional assumption about the generator states $| \Phi_{\qvec} \rangle$, known as the Gaussian overlap approximation (GOA)~\cite{brink1968,reinhard1987,krappe2012}. In its simplest formulation, the GOA assumes that the overlap between two generator states $\langle \Phi_\qvec | \Phi_\qpvec \rangle$ has a Gaussian shape that depends on the difference $(\qvec-\qpvec)$. In this work we use a more flexible version of this method introduced in~\cite{kamlah_derivation_1973,onishi1975,god1985-a}. We assume that one can find a change of variables $\qvec\rightarrow\boldsymbol{\alpha} = \boldsymbol{\alpha}(\qvec)$ such that the overlap reads
\begin{equation}
\label{eq:goa}
\langle \Phi_\qvec | \Phi_\qpvec \rangle \simeq
\operatorname{exp}\left( -\frac{1}{2} \sum_k \left[ \alpha_k(\qvec)- \alpha_k(\qpvec)\right]^2 \right).
\end{equation}
Within this approximation, the Hill-Wheeler equation reduces to a local, time-dependent Schr\"odinger-like equation in the space $\mathcal{Q}$ of the coordinates $\qvec$,
\begin{equation}
\label{eq:tdgcmgoa}
i\hbar \frac{\partial g(\qvec,t)}{\partial t} = \hat{H}_\text{coll}(\qvec) \, g(\qvec,t).
\end{equation}
The complex function $g(\qvec,t)$ is the unknown of the equation. It is related to the weight function $f(\qvec, t)$ appearing in (\ref{eq:gcmApprox}) and contains all the information about the dynamics of the system. The collective Hamiltonian $\hat{H}_\text{coll}(\qvec)$ is a local operator acting on $g(\qvec,t)$,
\begin{equation}
\label{eq:Hcoll}
\hat{H}_\text{coll}(\qvec)=
-\frac{\hbar ^2}{2\gamma^{1/2}(\qvec)} \sum_{ij} \frac{\partial}{\partial q_i} \gamma^{1/2}(\qvec) B_{ij}(\qvec) \frac{\partial}{\partial q_j}  +  V(\qvec),
\end{equation}
where
\begin{itemize}
\item The potential $V(\qvec)$ and the symmetric collective inertia tensor $\Bvec(\qvec) \equiv B_{ij}(\qvec)$ can be determined from the original nuclear Hamiltonian $\hat{H}$ and the generator states $|\Phi_{\qvec} \rangle $. These quantities reflect the potential and kinetic properties of the system in the collective space.
\item The metric $\gamma(\qvec)$ is a positive, real, scalar field introduced by the change of variable $\qvec\rightarrow\boldsymbol{\alpha}(\qvec)$ used for the GOA approximation, see Eqs.(\ref{eq:G})-(\ref{eq:Gdet}).
\end{itemize}
Equation~ (\ref{eq:tdgcmgoa}) will be referred to as the TDGCM+GOA equation.

Our approach to compute fission yields is essentially a two-step process. We first compute the static fields $V(\qvec)$, $\Bvec(\qvec)$ and $\gamma(\qvec)$ for a given range of the collective variables. Then we determine the dynamics of the system by numerically solving the TDGCM+GOA equation. Eventually, the fission yields are extracted as the flux of the collective wave packet $g(\qvec,t)$ through a hyper-surface of the collective space corresponding to scission configurations. In the following sections, we describe in more details each of these steps.


\subsection{Static part of the calculation}

The determination of the various fields in (\ref{eq:Hcoll}) is by far the most time- and resource-consuming part of the calculation. The size of the calculation grows exponentially with the number $N$ of collective variables needed to ensure a good fidelity of the physics model. It is well-known that in macroscopic-microscopic models, $N$ must be of the order of 5 in order to reach a satisfactory level of accuracy \cite{moeller2001}. Owing to the variational principle, it is often implicitly assumed that $N\leq 2$ should be sufficient in DFT calculations. Our analysis will show that $N=3$ collective variables are probably needed to make the leap in accuracy needed by applications.


\subsubsection{\label{sec:generatorStates} Determination of the generator states}

Generator states are obtained by solving the Hartree-Fock-Bogoliubov (HFB) equations of the EDF method under constraints on the expectation values of the axial quadrupole, $\hat{Q}_{20}$, and octupole, $\hat{Q}_{30}$, moments. To model the particle-hole channel we use either the SkM* parametrization of the Skyrme functional, or the D1S parametrization of the Gogny functional. In the case of the Skyrme EDF calculations, the pairing channel is modeled by a density-dependent surface-volume potential characterized by two pairing strengths, one for neutrons and one for protons. These pairing strengths are adjusted on the local three-point odd-even mass difference indicator in $^{240}$Pu; details are given in \cite{schunck2014}. As customary for calculations with the Gogny EDF, the same parametrization of the Gogny force is used in both particle-hole and particle-particle channels.

Before detailing the methods used to produce the complete self-consistent PESs, we briefly recall two important features extensively discussed in Ref.~\cite{dubray2012}. The first one is related to the fact that iterative HFB solvers may converge to a local minimum of the total binding energy instead of the targeted global minimum, which may lead to a spurious hysteresis behavior of the total binding energy when computing the PES iteratively from neighbor to neighbor. The second noteworthy characteristics of self-consistent PES is the presence of what will be referred to by the generic label of ``discontinuities''. Since each point of the PES is a solution to the variational problem, there is no guarantee that two neighboring points in the $(\hat{Q}_{20},\hat{Q}_{30})$ collective space, which is a two-dimensional projection of the full Hilbert space, are actually ``close'' in the full Hilbert space. As a consequence, the expectation value of various one body observables may not be a continuous function of the collective variables. 

Calculations with the Skyrme EDF are performed with the HFODD solver of \cite{schunck2012} in a one-center harmonic oscillator (HO) basis including up to $N_{\mathrm{max}} = 30$ HO shells. We recall that in HFODD basis states are characterized by the Cartesian frequencies $\hbar\omega_x$, $\hbar\omega_y$ and $\hbar\omega_z$; the equivalent spherical frequency $\omega_0 = (\omega_x\omega_y\omega_z)^{1/3}$ and the (here, axial) basis deformation $\omega_z/\omega_x$ are parametrized as a function of the quadrupole moment according to the formulas given in \cite{schunck2014}. In the pairing channel, the usual cut-off in the quasi-particle space is set at $E_\mathrm{cut} = 60$ MeV for calculations with surface-volume pairing. Each point in the PES is computed independently of one another using several successive preconditioners (solutions to a Woods-Saxon potential, axial calculation). 

Calculations with the Gogny EDF are performed with a HFB solver based on a two-center, cylindrical, HO basis \cite{berger_1984}, which is particularly well suited for the description of highly elongated systems. The basis includes states of both oscillators from all shells up to $N_{\mathrm{max}} = 11$ and the basis parameters are optimized at each deformation. The D1S PES was computed using a special iterative retro-propagation technique. For a given set of constraints, the HFB iterative solver is initialized from a previously converged HFB solution with neighboring constraints. If a discontinuity is found between two neighboring pre-scission HFB solutions, the solution with the highest total binding energy is recalculated, starting from the one with the lowest total binding energy. This algorithm can be automated and provides a reproducible method to generate the PES while removing hysteresis. Within the domain of pre-scission configurations, this approach ``explores'' all discovered valleys and selects among them the ones with the lowest energy. Points in the fusion valley (post-scission) are not retro-propagated so that the GCM state (\ref{eq:gcmApprox}) is enriched with a maximum number of pre-scissionned configurations.
To conclude with, note that both methods described above provide PES that may contain discontinuities. This feature is an intrinsic limitation of our 2D self-consistent description. 


\subsubsection{\label{subsubsec:collectiveFields} Collective Fields}

The collective inertia tensor is an essential input to the collective Hamiltonian (\ref{eq:Hcoll}). In a strict TDGCM approach to nuclear dynamics, the inertia tensor is fully derived from the many-body Schr\"odinger equation alongside the collective equation, yielding what is referred to in the literature as the GCM inertia tensor (or GCM collective mass) \cite{ring2000}. However, earlier studies based on translational invariance pointed out that the GCM inertia is not a satisfactory approximation of the nuclear collective inertia \cite{peierls1962}. The most popular alternative for the collective inertia tensor is to use the expression obtained by taking the adiabatic limit of the time-dependent Hartree-Fock-Bogoliubov equations. This ATDHFB inertia tensor was shown to yield the exact translational mass \cite{baranger1978}. It is important to make the following three remarks:
\begin{itemize}
\item If the collective potential $V(\qvec)$ is taken as the HFB potential energy, the quantization of the ATDHFB equations yields the same expression as Eq.(\ref{eq:tdgcmgoa})-(\ref{eq:Hcoll}). It is then fully justified to use either the GCM or the ATDHFB inertia tensor. In the GCM framework, however, the derivation of the collective Schr\"odinger equation yields additional zero-point energy (ZPE) corrections to the HFB potential energy. These corrections can be expressed as a function of the mass tensor. It has thus argued that one could postulate a phenomenological ZPE for the quantized ATDHFB equation of motion by simply taking the GCM ZPE formula and substituting the ATDHFB inertia tensor in it.
\item Both the GCM and ATDHFB inertia are formulated as a function of the inverse of a matrix analog to the QRPA matrix \cite{schunck2016}. In practice, inverting such a matrix is a tremendous challenge, and nearly all applications have been obtained in the cranking approximation, where the residual interaction among quasi-particles is neglected and the QRPA matrix is diagonal and given by $\mathcal{M}_{ij,\mu\nu} = (E_{i} + E_{j})\delta_{i\mu}\delta_{j\nu}$.
\item Irrespective of the prescription retained for the collective inertia tensor, its expression also involves the derivatives of the generalized density with respect to the collective variables. In practice, these derivatives are often approximated locally at point $\qvec$ by introducing a quasi-particle representation of the collective momentum operator defined by $\gras{P} = (\hat{P}_{1},\dots,\hat{P}_{N})$ with $\hat{P}_{k} = -i\hbar\partial/\partial q_{k}$. When combined with the aforementioned cranking approximation, this approach has been dubbed the perturbative cranking \cite{baran2011}. Note that it can be applied both for the GCM and for the ATDHFB inertia tensor.
\end{itemize}
In this work, we have employed the perturbative cranking approximation of the inertia tensor, using either the GCM or ATDHFB formula.

In the GCM+GOA approximation, the collective inertia tensor $\Bvec$, which is the inverse of the collective mass tensor $\Mvec$, reads
\begin{multline}
B_{ij}(\qvec) = [M^{-1}]_{ij}(\qvec) \\
 = \frac{1}{2\hbar^2} \sum_{kl}
 G^{-1}_{ik}(\qvec)
 \left(
 \left.\frac{\partial ^2 h(\avec, \apvec)}{\partial a_k \partial a_l'}\right|_{\qvec}
 - 
 \left.\frac{\partial ^2 h(\avec, \apvec)}{\partial a_k \partial a_l}\right|_{\qvec} 
 \right.\\
 + \displaystyle
 \left.
 \left. \sum_{n}\Gamma^n_{kl}(\avec) \frac{\partial h(\avec, \apvec)}{\partial a_n}\right|_{\qvec}
 \right)
 G^{-1}_{lj}(\qvec).
\label{eq:B_general}
\end{multline}
In this expression, all derivatives are evaluated at point $\avec=\apvec=\qvec$. The reduced Hamiltonian $h(\avec,\apvec)$ is defined from the norm and Hamiltonian kernels by
\begin{equation}
h(\avec,\apvec) 
= \frac{\mathcal{H}(\avec,\apvec)}{\mathcal{N}(\avec,\apvec)} 
=  \frac{\langle \Phi_{\avec} |\hat{H} | \Phi_{\apvec} \rangle}{\langle \Phi_{\avec}| \Phi_{\apvec} \rangle},
\end{equation}
and the metric tensor $\Gvec \equiv G_{ij}$, which characterizes the change of variable $\qvec\rightarrow\gras{\alpha}(\qvec)$, is defined by
\begin{equation}
G_{ij}(\qvec)= \sum_k \frac{\partial \alpha_k}{\partial q_i}\frac{\partial \alpha_k}{\partial q_j}.
\label{eq:G}
\end{equation}
The determinant of this matrix gives the metric $\gamma(\qvec)$ in the TDGCM+GOA equation,
\begin{equation}
\gamma(\qvec)= \text{det}\; \Gvec(\qvec).
\label{eq:Gdet}
\end{equation}
The notation $\Gamma^n_{kl}$ stands for the Christoffel symbol. It is related to the metric tensor $\Gvec(\qvec)$ through the relation
\begin{equation}
\Gamma^n_{kl}(\qvec)= \frac{1}{2} \sum_i G^{-1}_{ni}(\qvec)
\left(
\frac{\partial G_{ki}}{\partial a_l} + \frac{\partial G_{il}}{\partial a_k} - \frac{\partial G_{lk}}{\partial a_i}
\right).
\end{equation}
The last term in the expression of the collective inertia tensor comes from the dependency of the metric tensor $\Gvec$ on the collective variables $\qvec$. This term is not present in the ``constant width'' version of the GCM+GOA.

The potential energy term $V(\qvec)$ is the sum of the HFB energy at point $\qvec$ and zero-point energy corrections,
\begin{multline}
V(\qvec)= \langle \Phi_{\qvec} | \hat{H} | \Phi_{\qvec} \rangle - \frac{1}{2} \text{Tr}\; \Bvec\Gvec \\
- \frac{1}{8} \text{Tr}\left[ \Gvec^{-1} \frac{\partial^2 \langle \Phi_{\qvec} | \hat{H} | \Phi_{\qvec} \rangle }{\partial q_{i}\partial q_{j}}  \right].
\label{eq:Vcoll_general}
\end{multline}
The potential is formally the same as in the constant width approximation, but the $\Gvec$ inertia tensor is now space-dependent.

Equations (\ref{eq:B_general}) and (\ref{eq:Vcoll_general}) can be simplified in the particular case of the perturbative cranking approximation. In this case, one introduces the moments 
\begin{equation}
M_{ab}^{(K)} = \frac{1}{2} ( Q_{a}^{12\,*},\ Q_{a}^{12}) \mathcal{M}^{-K} \left(\begin{array}{c} Q_{b}^{12} \\ Q_{b}^{12\,*} \end{array}\right),
\label{eq:moments}
\end{equation}
associated with the constraints $q_{a}$ and $q_{b}$. Equation (\ref{eq:moments}) is written in the quasi-particle basis. The matrix $\mathcal{M}^{-K}$ is the inverse of the $K^{\text{th}}$ power of the QRPA matrix. In the cranking approximation, it is simply given by $(\mathcal{M}^{-K})_{ij,\mu\nu} = \delta_{i\mu}\delta_{j\nu}/(E_{i}+E_{j})^{K}$. The matrix of these constraint operators have the generic block structure
\begin{equation}
\tilde{\mathcal{Q}}_{a} = \left(\begin{array}{cc}
Q_{a}^{11} & Q_{a}^{12} \\
Q_{a}^{21} & Q_{a}^{22}
\end{array}\right).
\end{equation}
The matrix element $M_{ab}^{(K)}$ is obtained by linearizing each of the block matrices involved and taking a regular scalar product. Based on this definition, one can show that the metric tensor $\Gvec$ reads
\begin{equation}
\Gvec = \frac{1}{2} [\Mvec^{(1)}]^{-1} \Mvec^{(2)} [\Mvec^{(1)}]^{-1},
\label{eq:G_local}
\end{equation}
and the collective inertia tensor becomes
\begin{equation}
\Bvec = \frac{1}{4} \Gvec^{-1} [\Mvec^{(1)}]^{-1} \Gvec^{-1}.
\label{eq:B_local}
\end{equation}
Equations (\ref{eq:G_local})-(\ref{eq:B_local}) only depend on the characteristics of the constrained HFB solutions at point $\qvec$. 


\subsection{\label{subsec:dynamics} Dynamic part of the calculation}

The main advantage of the adiabatic approach to fission is the perfect decoupling between static and dynamical aspects of the process. Once the potential energy surface $V(\qvec)$ has been computed and the collective inertia $\Bvec$ and metric tensor $\Gvec$ are determined, the time-evolution can be performed independently. In this section, we discuss practical aspects of the time evolution, including inputs and implementation.


\subsubsection{\label{subsubsec:initialState} Initial state}

The starting point of the dynamics calculation is a collective wave packet $g(\qvec,t=0)$ representing the compound nucleus after the absorption of a low energy neutron. In the full scattering theory picture, such a state depends both on the structure of the compound system and on the neutron entrance channel. In this work, the initial state is built arbitrarily from the assumption that it should mainly be localized at low deformations, in the inner potential well of the potential energy surface. To simulate such an induced fission event, we also impose the average energy of the initial state to be 1 MeV higher than the inner fission barrier. These requirements are not enough to uniquely define the initial state but they may constrain strongly the resulting fission yields.

To build a low deformation wave packet, we begin with computing a series of quasi-bound states as described in~\cite{goutte2005,younes2012-a}. This is achieved by extrapolating the inner potential barrier with a quadratic form, which defines a new potential $V'(\qvec)$. Replacing the original potential $V(\qvec)$ by its extrapolated counterpart defines a new collective Hamiltonian $\hat{H}'_\text{coll}(\qvec)$. Its eigenvalues are the quasi-bound states $\{g_{k}(\qvec)\}$,
\begin{equation}
\label{eq:eigenStates}
\hat{H}'_\text{coll}(\qvec) \, g_{k}(\qvec)= E_k \, g_{k}(\qvec).
\end{equation}
The normalized solutions of Eq.~(\ref{eq:eigenStates}) are defined up to a phase. Since the collective Hamiltonian $H'(\qvec)$ is real, we choose real solutions $g_k$ with the following sign convention,
\begin{equation}
 \int_{\qvec\in\{Q_{20}>0,\,Q_{30}>0\}} g_k(\qvec) d\qvec>0.
\end{equation}
The eigenstates $g_{k}(\qvec)$ with the lowest energies are all localized at low deformation and contain different phonons in the $\hat{Q}_{20}$ and/or $\hat{Q}_{30}$ degrees of freedom. In practice, we determine the first 100 of them using a Krylov-Schur algorithm implemented in the SLEPc library~\cite{hernandez_slepc:_2005}. The highest energy state lies roughly 16 MeV above the inner barrier energy.

In order to investigate the sensitivity of our calculations to different initial conditions, we have solved the TDGCM+GOA equations for two different choices of initial states, both based on the set of quasi-bound states $g_k$. Our first choice relies only on the collective ground-state $g_0$. The modulus of $g_0$ is roughly a Gaussian centered on the minimum of $V'(\qvec)$ and characterized by a width close to the dimension of the first potential well. Since its average energy lies below the fission barrier, we boost this state in the $Q_{20}>0$ direction in order to simulate a fission event,
\begin{equation}
g(\qvec, t=0) = g_0(\qvec)\operatorname{exp}(i k Q_{20}).
\end{equation}
The amplitude $k$ of the boost is determined so that the average energy of the initial state lie 1 MeV above the inner fission barrier ($B_{I}$).

Our second choice consists in building the initial state as a Gaussian superposition of the quasi-bound states $g_k$,
\begin{equation}
g(\qvec, t=0) = \sum_k \operatorname{exp}\left(\frac{ (E_k- \bar{E})^2}{2 \sigma^2}\right) g_k(\qvec).
\end{equation}
Given $\sigma$, the parameter $\bar{E}$ is chosen so that the average energy of the initial state is again 1 MeV above the inner fission barrier. The parameter $\sigma$ controls the energy spread of this wave packet. Its value is set to 0.5 MeV so that the weight associated to $g_k$ is significant ($>5\%$) in the range $[B_{I}; B_{I} + 2 \text{ MeV}]$.


\subsubsection{Boundary conditions}

Practical as well as physical reasons, discussed further in Sec.~\ref{sec:frontierPosition}, justify solving the TDGCM+GOA equation in a finite domain $\Omega$ of the $(\hat{Q}_{20},\hat{Q}_{30})$ collective space. However, as time goes by, the initial wave packet propagates and should eventually leak outside of the simulation box. To account for this effect, $\Omega$ is extrapolated with a band $\mathcal{B}$ of artificial points continuously connected to its boundary. In this new region, the potential decreases linearly as a function of the distance to $\Omega$, whereas the inertia tensor and the metric are kept constant.
On top of this, an imaginary term $-i\hbar A(\qvec)$ (with $A(\qvec)$ a real scalar field) is added to the collective Hamiltonian. This mechanism absorbs progressively the leaking wave packet and avoids spurious reflections. We define $A(\qvec)$ as a simple third order polynomial that increases smoothly from 0 on the inner border of the band $\mathcal{B}$ and reaches its maximum at the outer border.


\subsubsection{Collective dynamics with the FELIX solver}

Starting from an initial state as described in Sec.\ref{subsubsec:initialState}, we solve the TDGCM+GOA equation with the finite element solver FELIX. A first version of this code was previously released under the GPL-2 open source license and is available in the Computer Physics Communication Library~\cite{regnier2016}. For the purpose of this work, we developed a new version that takes into account the metric term $\gamma(\qvec)$ in Eq.~(\ref{eq:tdgcmgoa}). This section briefly recalls the numerical implementation of the TDGCM+GOA equation in FELIX and emphasizes the recent updates.

The solver relies on a continuous Galerkin finite element method to discretize the space $\mathcal{Q}$ of the collective coordinates. The first step of this method consists in partitioning the domain of interest into a mesh of 2-dimensional simplices (triangles). The numerical solution is then expanded on the basis of the continuous Lagrange elements of order 2 denoted here as $\{l_i(\qvec)\}_i$,
\begin{equation}
\label{eq:gExpand}
g(\qvec,t)= \sum_{ i=1}^m G_i(t) \, l_i(\qvec).
\end{equation}
Applying the Galerkin finite element method, we search for a numerical solution $g(\qvec,t)$ of the form (\ref{eq:gExpand}) that verifies
\begin{multline}
\forall t \in [0,t_{\text{max}}], \forall i \in [1,m]: \quad \\
\langle \l_i(\qvec) | \left[i \hbar \frac{\partial}{\partial t} - \hat{H}_\text{coll}(\qvec)\right] | \, g(\qvec,t)  \rangle = 0,
\end{multline}
where the scalar product $\langle . |  . \rangle$ includes the metric $\gamma(\qvec)$ as follows,
\begin{equation}
\langle f | g \rangle = \int_\Omega d\qvec\; f^*(\qvec) \,g(\qvec) \gamma^{1/2}(\qvec).
\end{equation}
This process yields a discretized system of $m$ equations with the $m$ coefficients $G_i(t)$ as unknowns. It can be written in the condensed matrix form
\begin{equation}
\label{eq:spaceDiscretizedEvolution}
i\hbar \mathbf{M} \, \frac{\partial \mathbf{G}(t)}{\partial t}= [\mathbf{H} -i\hbar \mathbf{A}]\mathbf{G}(t),
\end{equation}
where $\mathbf{G}(t)$ denotes the $m$-dimensional vector of coefficients. The $m\times m$ real matrices $\mathbf{M}, \mathbf{H}$ and $\mathbf{A}$ are defined by
\begin{equation}
\begin{array}{l}
M_{ab}= \langle l_a(\qvec) | l_b(\qvec) \rangle, \medskip\\ 
A_{ab}= \langle l_a(\qvec) | A(\qvec)l_b(\qvec)\rangle, \medskip\\ 
H_{ab}= \langle l_a(\qvec) | H_\text{coll}(\qvec)l_b(\qvec)\rangle.
\end{array}
\end{equation}
Once the matrix elements are computed, FELIX performs the time integration of equation (\ref{eq:spaceDiscretizedEvolution}) based on a Crank-Nicolson scheme. This unitary and implicit method requires solving a $m\times m$ complex sparse linear system at each time step. Our  implementation of sparse matrix inversions is based on the Quasi Minimal Residual method as described in Ref.~\cite{freund_conjugate_1992}.


\subsubsection{\label{sec:fragmentDistribution} Fission fragment distributions}

A fully quantum mechanical derivation of the properties of the fission fragments from TDGCM states is faced with several major difficulties. The first challenge is to describe two well separated fragments from a single density associated with the compound system. The second challenge is to project each of the fragment wave functions onto the eigenspace of the observable of interest ($Z$, $N$, angular momentum, etc.)

In this work, we adopted the empirical approach described in \cite{berger_time-dependent_1991,younes2012-a}, which is based on computing the flux of the collective wave packet. The TDGCM+GOA equation implies the following continuity equation for the quantity $|g(\qvec, t)|^2 \gamma^{1/2}(\qvec)$,
\begin{equation}
\frac{\partial}{\partial t} |g(\qvec, t)|^2 \gamma^{1/2}(\qvec) = -\nabla \cdot \Jvec(\qvec, t),
\end{equation}
where the current $\Jvec(\qvec, t)$ reads
\begin{multline}
\Jvec(\qvec,t)=
\frac{\hbar}{2i} \gamma^{1/2}(\qvec) \Bvec(\qvec) \\
\times \big[ g^*(\qvec,t) \nabla g(\qvec,t) - g(\qvec,t) \nabla g^*(\qvec,t) \big].
\end{multline}
As time goes by, the collective wave packet progressively escapes the simulation domain $\Omega$ through its boundary (also denoted frontier in this paper). Each infinitesimal surface element of the frontier line is assumed to be the entrance point of a channel associated with a given fragmentation $(A_H,A_L)$. The probability to measure this fragmentation is estimated from the time-integrated flux $F(\xi,t)$ through each of the oriented surface elements $\xi$ of the frontier line on the discretized mesh,
\begin{equation}
F(\xi,t) = \int_{t=0}^{t} dt \int_{\qvec\in\xi} \Jvec(\qvec,t)\cdot d\Svec.
\label{eq:fluxDef}
\end{equation}
The fission fragment mass yield for mass $A$ is defined formally as
\begin{equation}
Y(A) \propto \sum_{\xi\in\mathcal{A}} \lim_{t\rightarrow +\infty} F(\xi, t),
\label{eq:yield}
\end{equation}
where $\mathcal{A}$ is the set of all oriented hyper-surfaces $\xi$ belonging to the frontier hyper-surface such that one of the fragments has mass $A$. In practice, the fragment mass number at any point of the frontier is calculated as the integral of the nuclear density on one side of the neck \cite{schunck2014}. This procedure produces non integer values. Moreover, one elementary surface $\xi$ may contain several fragmentations, especially if the mesh is coarse. For these reasons, we equally distribute the flux component $F(\xi,t)$ between the masses calculated at the vertices of the elementary surface $\xi$,
\begin{equation}
Y(A) = C \sum_{\xi} \frac{1}{N} \sum_{v\in\mathcal{A(\xi)}} \lim_{t\rightarrow +\infty} F(\xi, t).
\label{eq:practicalYield}
\end{equation}
The sum on $\xi$ runs over the whole scission hyper-surface. The set $\mathcal{A(\xi)}$ contains the vertices of $\xi$ where the mass of one of the fragments is in the interval $[A-1/2; A+1/2]$. The normalization constant $C$ is chosen as usual such that 
\begin{equation}
\sum_{A=0}^{A_{\text{total}}} Y(A) = 200.
\label{eq:yield_norm}
\end{equation}

Albeit simple to implement and containing sufficient physics, we note that such an evaluation of the number of particles in the fragments is not performed in a fully quantum mechanical formalism. Several effects should be taken into account, which could ultimately impact our predictions:
\begin{enumerate}
\item First, the HFB solutions of the fissioning system with mass $A$, in particular at scission, do not have good numbers of protons and neutrons. As a result, the HFB wave function contains components with particle number $A\pm 2$, $A\pm 4$, etc. As a result, the yield of each fragment includes a spurious contribution coming from the split of one of these subsystems. We thus expect that projecting the HFB wave function of the compound nucleus into good particle numbers may reduce the width of the fission yields.
\item Eventually, the number of particles in the fragments themselves should also be computed by projecting the full wave-function on good particle number. Each point of the frontier line would therefore be associated with a distribution of fragment mass and charges \cite{scamps2015}. Taking into account this distribution would probably have the opposite effect of the projection on good particle number for the fissioning nucleus and should broaden the fission yields.
\item At scission, the fragments are still strongly entangled. This point was first mentioned in \cite{younes2011}, where a technique to disentangle the fragments based on a unitary transformation of the quasi-particles was introduced; see also \cite{schunck2014} for details. If this operation is not performed, we can expect exchanges of nucleons between the two pre-fragments and therefore a broadening of the raw fission yields.
\end{enumerate}
Although difficult to quantify at this point, we expect that the cumulative effect of this missing physics will induce a global widening of the fission yields. In this work, we chose to account for such effects phenomenologically by convoluting the raw flux with a Gaussian function of the number of particles. The width of this Gaussian is set to 4 for the mass yields; see also discussion in Sec.\ref{sec:frontierPosition}. As the charge to mass ratio is roughly 2.55 the width for the number of charge is set to $4/2.55=1.6$. We briefly discuss in section \ref{subsec:gaussian} the uncertainty coming from the choice of this Gaussian width.


\section{\label{sec:results} Results}

In this section, we present the result of our simulations. We discuss the characteristics of the potential energy surface depending on the EDF used, the numerical convergence of the TDGCM+GOA calculations and the sensitivity of the yields to various inputs of the calculation, such as the EDF, the initial state, the collective inertia tensor and the convolution width.

\begin{figure*}[!ht]
\includegraphics[height=9cm]{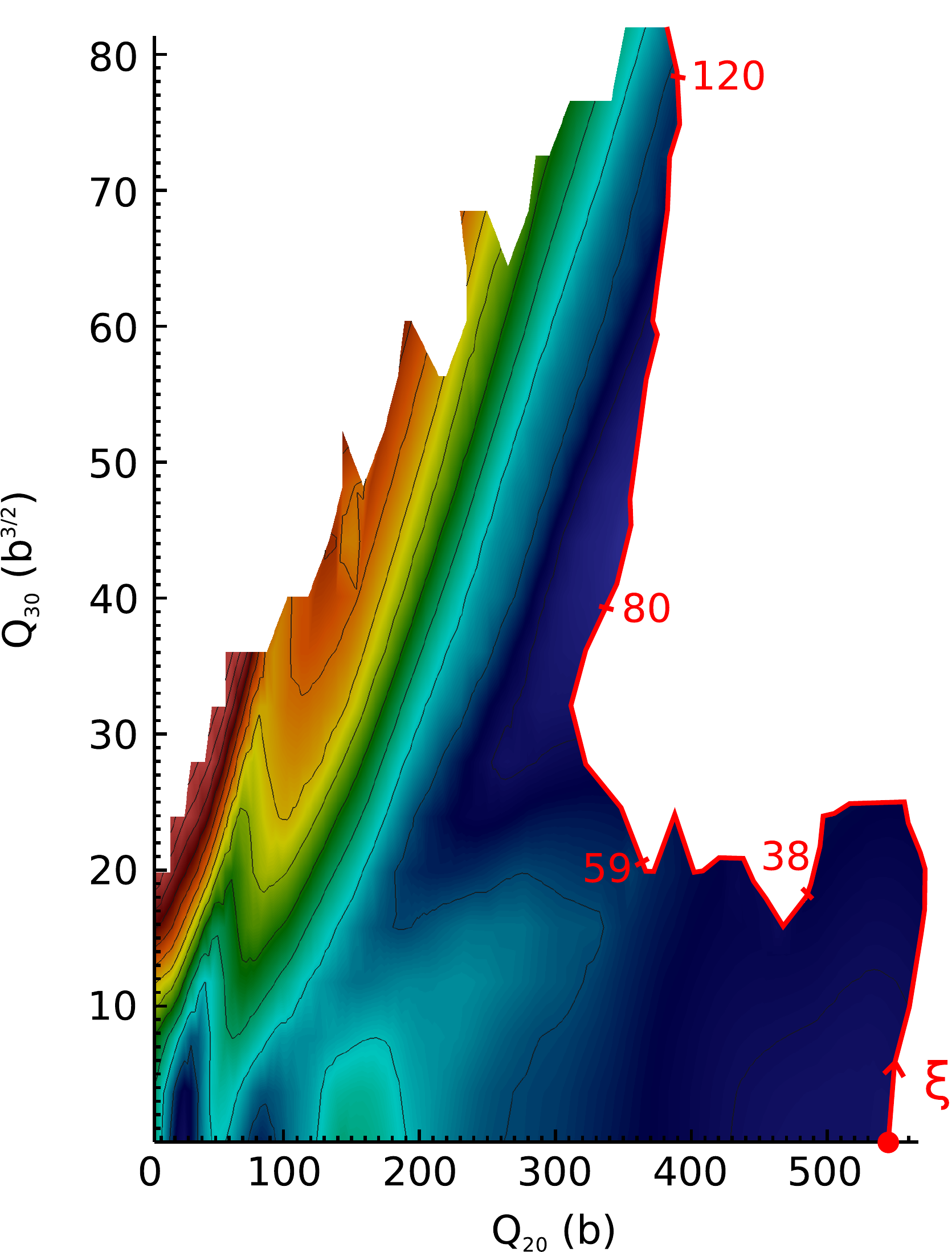}
\includegraphics[height=8.02cm]{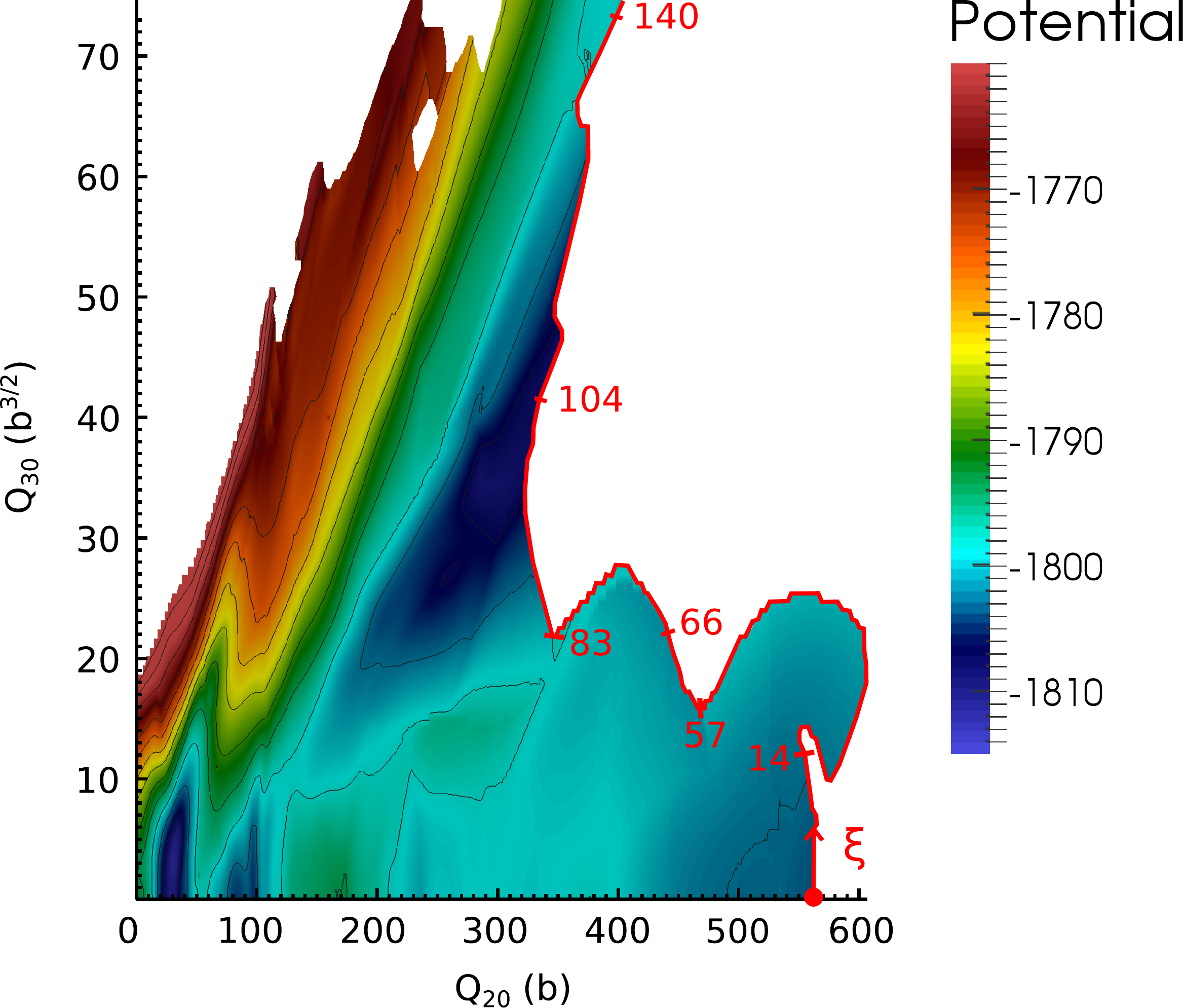}
\caption{(color online) Potential energy surfaces obtained with the SkM* (left) and D1S (right) EDF in axial symmetry. The red line separates configurations with $Q_{N}>4$ from the others. The curvilinear abscissa $\xi$ starts at the symmetric scission points and runs along the frontier (red line). Values of $\xi$ are indicated along the scission line.}
\label{fig:pes}
\end{figure*}


\subsection{\label{sec:staticResults}  Collective landscape in the $\gras{(\hat{Q}_{20},\hat{Q}_{30})}$ space}

As mentioned in section \ref{subsec:dynamics}, the first step toward the dynamical description of fission is to determine the characteristics of the potential energy surface of the compound nucleus in the collective space, here defined by the two collective variables $(\hat{Q}_{20},\hat{Q}_{30})$. In terms of the cylindrical coordinates $(z, \rho, \theta)$, our choice for the quadrupole and octupole moments is
\begin{equation}
\begin{array}{l}
 \displaystyle\hat{Q}_{20}(z,\rho,\theta) = 2 z^2 - \rho^2 \\
 \displaystyle\hat{Q}_{30}(z,\rho,\theta) = \sqrt{\frac{7}{4\pi}}(z^3 -\frac{3}{2} z \rho^2)
\end{array}
\end{equation}
With this convention, we computed the potential energy surface for the Gogny D1S and the Skyrme SkM*~\cite{bartel1982} EDF. The range of collective variables is 0 to 600 b for $\hat{Q}_{20}$ and from 0 to 80 b$^{3/2}$ for $\hat{Q}_{30}$. The D1S potential energy surface contains roughly 37,000 un-scissioned points distributed on a regular mesh of step $\Delta_{q_{20}}=1$ b, $\Delta_{q_{30}}=0.75$ b$^{3/2}$. Because the HFODD DFT solver used for  SkM* calculations breaks axial symmetry, the time of calculation is significantly longer. As a result, fewer points were computed and the unsciscionned SkM* PES grid is coarser, with approximately 1,500 points. Most of the points are nodes of a regular grid with $\Delta_{q_{20}}=5$ b, $\Delta_{q_{30}}=4$ b$^{3/2}$. To compensate for this coarse grid, additional calculations were performed in the scission region, where system properties vary quickly. In order to ensure that the characteristics of the mesh do not affect our conclusions, we built an auxiliary potential energy surface for D1S based on 2,500 points sampled from the original 37,000. These points are distributed in the same way as for the SkM* mesh. We checked that fission fragment distributions obtained with this coarser grid is essentially unchanged compared to the original one.

Figure~\ref{fig:pes} shows the potential energy surface obtained. Note that these two PES contain a zero-point energy correction, i.e. the potential energy corresponds to (\ref{eq:Vcoll_general}) with the GCM inertia. Keeping in mind the comments made in \ref{subsubsec:collectiveFields}, this means that we work here in a strict GCM framework. Both interactions yield a similar overall topology, although the potential variations are often more pronounced with D1S. We clearly distinguish an inner symmetric fission barrier followed by an outer asymmetric one. The table~\ref{tab:barriers} lists the position and height of these fission barriers. Both EDF predict the same positions for the saddle points, whereas the barrier heights may vary by more than 3 MeV. Note that calculations with the Gogny force were performed with an axial code; including triaxiality effects would roughly lower the barriers by 3 MeV for the D1S case \cite{delaroche2006}. In the rest of this paper, and for the sake of comparison between Skyrme and Gogny EDF, all results of TDGCM simulations rely on axial PES unless explicitly specified.

\begin{table}[!ht]
\caption{(color online) Characteristics of the PES. The ground-state energy ($E_0$) in the first potential well is given in MeV. The height of the inner and outer fission barriers (B$_I$/B$_{II}$) are in MeV relative to $E_0$.}
\begin{ruledtabular}
\begin{tabular}{cccccc}
         & $E_0$       & B$_I$        & B$_{II}$ & $Q^{II}_{20}$ (b) & $Q^{II}_{30}$ (b$^{3/2}$) \\
SkM* (axial)  & -1808.89 & 9.3  & 7.6  & 128 & 9.5 \\
SkM* (triax.) & -1808.81 & 7.7  & 7.6  & 128 & 9.5 \\
D1S  (axial)  & -1810.81 & 12.4 & 11.9  & 130 & 10.4 \\
\end{tabular}
\end{ruledtabular}
\label{tab:barriers}
\end{table}


\subsection{ \label{sec:frontierPosition} Position of the frontier}

While static properties are computed over a wide range of the collective space, only a sub-domain $\Omega$ of all such HFB states is needed to define the GCM state (\ref{eq:gcmApprox}). The choice of $\Omega$ is an important step in our approach as the frontier of $\partial \Omega$ determines the output channels of the dynamics. Three criteria guided our selection:
\begin{enumerate}
\item Configurations with a potential energy much larger than the energy of the fissioning system do not contribute to the dynamics. As a consequence, all points with a potential energy greater than 100 MeV above the ground state are not included in $\Omega$.
\item The collective wave packet should follow continuous collective paths toward the frontier. In terms of dynamics, a wave packet passing through a discontinuity of the HFB states travels suddenly from one valley of the full space to another one. By doing so, it misses intermediate configurations and their associated potential barrier. Such spurious crossings are not physical and particular attention must be paid so that they do not occur inside the selected domain $\Omega$.
\item The frontier should contain configurations that are as close as possible to two well separated fragments. In our approach, each infinitesimal element of the frontier is considered an output channel. Therefore, the less entangled the fragments on the frontier are, the closer the output channels of the dynamics calculation will correspond to the actual output channels of fission.
\end{enumerate}

When describing fission in the $(\hat{Q}_{20},\hat{Q}_{30})$ collective space, scission is characterized by a discontinuity between two domains containing the pre- and post-scissioned configurations. As a consequence of the remark 2, the domain $\Omega$ should not include post-scissioned configurations. A relevant degree of freedom to describe scission is the Gaussian neck operator $\hat{Q}_N$ related to the number of particles in the neck. As emphasized in the work of Younes~\etal~\cite{younes2009-a}, this quantity shows a pronounced gap between pre- and post-scission configurations. Based on this observation, a possible choice for the domain $\Omega$ consists in keeping only the configurations with $Q_N>1$ (\textit{e.g.} pre-scissioned configurations). 

\begin{figure}[!ht]
\includegraphics[width=0.45\textwidth]{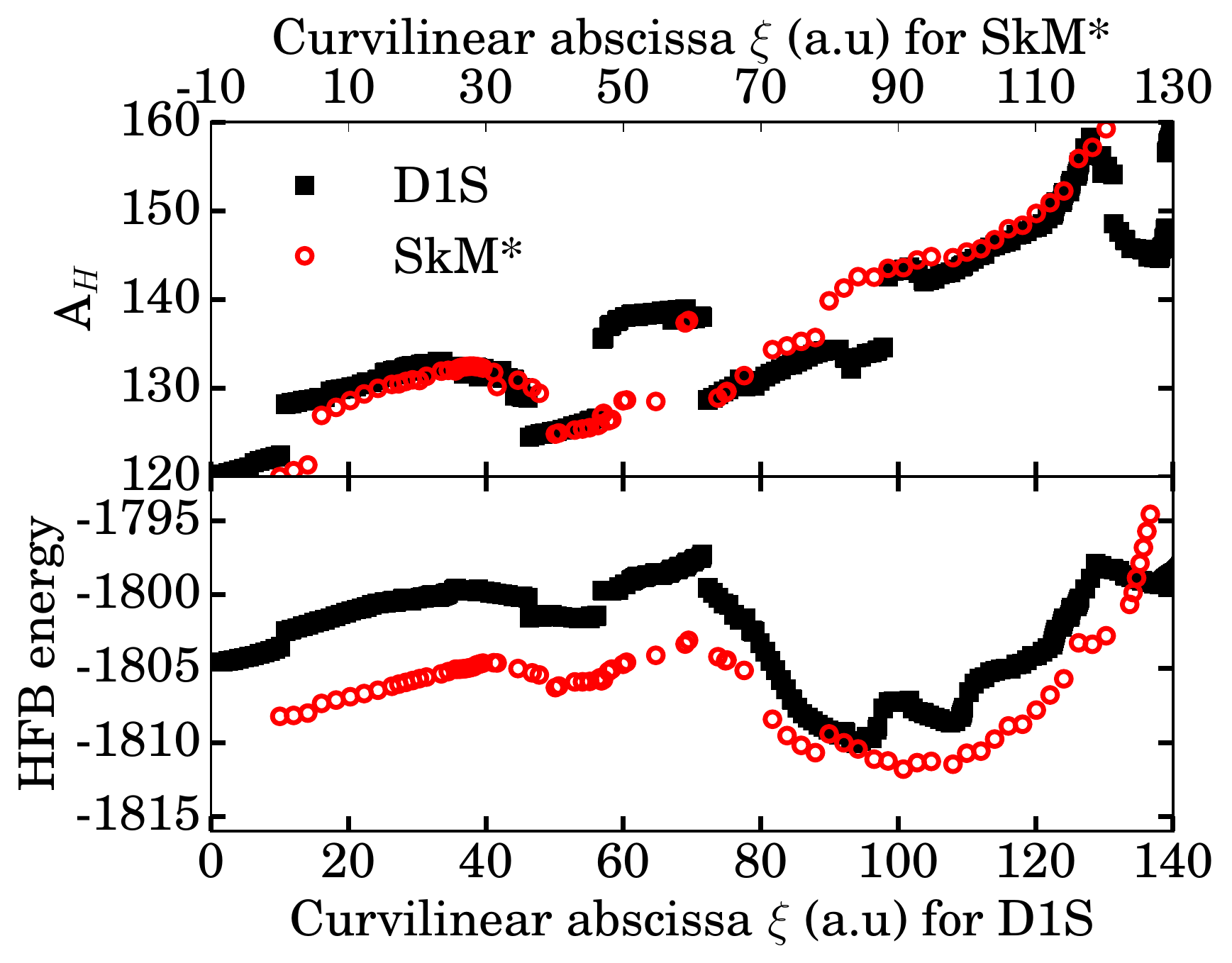}
\caption{(color online) Static properties of the system along the frontier of the domain defined by $Q_N > 1$. The position on the frontier is labeled by the curvilinear abscissa $\xi$ starting from the symmetric point ($Q_{30}=0$). The  SkM* abscissa is shifted by 10 arbitrary units so that the main structures match D1S. The HFB energy is expressed in MeV.}
\label{fig:massDiscontinuity}
\end{figure}

Figure~\ref{fig:massDiscontinuity} shows the evolution of the total energy and the heavy fragment mass number along the boundary of such a domain. The total binding energy obtained with D1S possesses several jumps ($\xi= 11, 47, 57, 72$) that may look incompatible with the retro-propagation algorithm described in Sec.~\ref{sec:generatorStates}. These singularities can actually be explained by the proximity of the fusion valley. Let us note $M$ and $M'$ two neighboring points on the frontier with different HFB energies. It turns out that a HFB calculation of the state $M'$ initialized with the converged state $M$ leads to a post-scission configuration and \textit{vice versa}. Such rare patterns in the total binding energy are therefore not related to possible remaining spurious hysteresis features in the PES. The mass of the heavy fragments is itself subject to several sharp variations as a function of the curvilinear abscissa $\xi$ characterizing the frontier. This behavior is the typical trademark of various discontinuities crossing the frontier.

\begin{figure}[!ht]
\includegraphics[height=7cm]{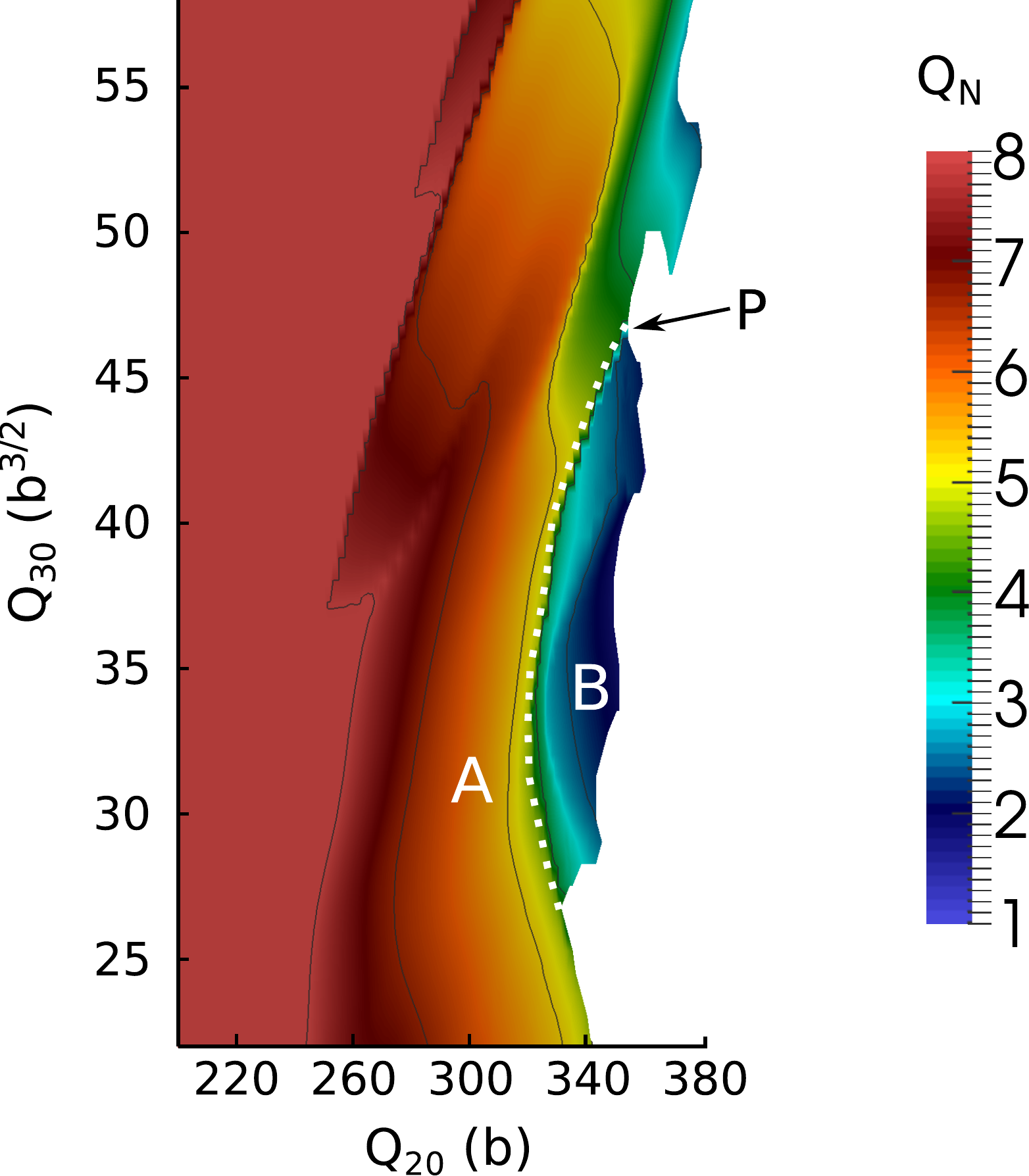}
\caption{(color online) Gaussian neck operator values in the vicinity of the exit of the asymmetric fission valley of the D1S PES. Valley A corresponds to the main asymmetric valley whereas region B corresponds to a different valley of the full Hilbert space projected in between the post-scission configurations and the valley A. The point P corresponds to abscissa $\xi\simeq100$ on the frontier defined by $Q_N>1$.}
\label{fig:fissionValley}
\end{figure}

\begin{figure}[!ht]
\includegraphics[width=0.45\textwidth]{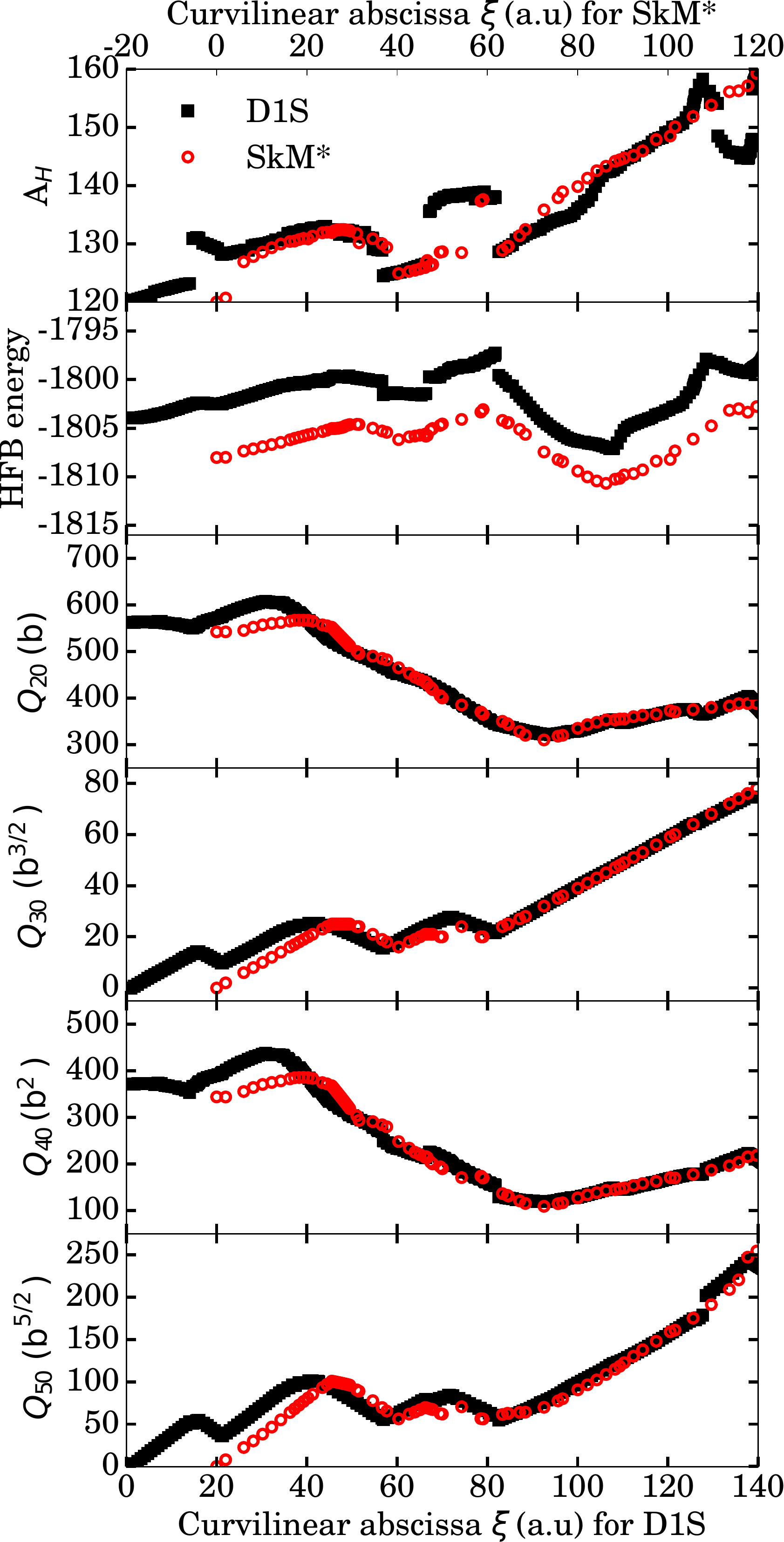}
\caption{(color online) Static properties of the system along a frontier of the domain defined by $Q_N > 4$. The position on the frontier is labeled by the curvilinear abscissa $\xi$ starting from the symmetric point ($Q_{30}=0$). The  SkM* abscissa is shifted by 20 arbitrary units so that the main structures match. The HFB energy is expressed in MeV and the multipole moments in powers of b.}
\label{fig:frontierQn4}
\end{figure}

To better highlight this feature, we discuss in more details the discontinuity present at $\xi \simeq 100$ in the D1S calculation. This discontinuity matters particularly as it is located in the midst of the asymmetric fission valley, where most of the wave packet leaks out of the domain $\Omega$.
Figure~\ref{fig:fissionValley} shows the expectation value of the $\hat{Q}_N$ operator in this area of the potential energy surface. We observe two distinct regions. Zone A corresponds to the main asymmetric valley. It can be continuously connected to the first potential well and contains configurations with a large neck ($Q_N > 5$). On the other hand, region B corresponds to the projection of a different valley of the full Hilbert space lying between post-scission configurations and the main asymmetric valley. Its HFB states are characterized by thinner necks with $Q_N$ ranging from 2 to 3. The connection of regions A and B at point P in our two dimensional space induces the discontinuity in mass present on the frontier at $\xi \simeq 100$. Note that the spread of this intermediate valley just before scission is a consequence of the special retro-propagation algorithm explained in Sec.~\ref{sec:generatorStates}. It is not observed in the SkM* surface, which explains why the SkM* frontier does not possess a similar mass jump. As explained in comment 2, the propagation of the wave packet from region A to B is not physical. In order to keep the best description of the dynamics in a 2D representation, the region B should therefore be excluded from the domain $\Omega$. 

Figure~\ref{fig:massDiscontinuity} shows that other areas along the frontier are affected by similar spurious discontinuities. To tackle this issue, the brute force strategy consists in locating each discontinuity and removing one by one the regions where spurious crossings may happen. An alternative to this tedious method is to choose a sufficiently high threshold for the expectation value of the $\hat{Q}_N$ operator so that the most impacting discontinuities are not included in $\Omega$. Figure~\ref{fig:frontierQn4} shows the static properties of both SkM* and D1S configurations along the frontier defined by $Q_N>4$. Both frontiers follow a similar trajectory in our two dimensional space resulting in close values of $\hat{Q}_{20}$ and $\hat{Q}_{30}$ as a function of the curvilinear abscissa $\xi$. In addition, higher order multipole moments also display a similar behavior for the two EDF. As we can see in figure~\ref{fig:fissionValley}, the threshold $Q_N= 4$ is sufficient to remove the region B from the main asymmetric valley. Hence, the mass discontinuity previously present at $\xi \simeq 100$ is avoided. Other mass jumps still remain at lower asymmetry (\textit{e.g.} $\xi=70, 58, 45$). Albeit persistent, the related discontinuities are in areas of the PES where the outgoing flux represents only a few percents of the outgoing wave packet. Consequently, their impact on the resulting fission yields is marginal in the context of our current applications. In conclusion, the criteria $Q_N> 4$ provides an operating definition of the domain $\Omega$ which avoids the main issues related to the discontinuities in our 2D collective space. Although more advanced criteria may be explored in future work, dynamical calculations presented in this paper rely on this threshold for $Q_{N}$.


\subsection{Numerical convergence of the dynamics}

The implementation of the TDGCM+GOA equation in FELIX depends on four parameters that control the numerical precision of the wave-packet propagation:
\begin{itemize}
\item The ``dynamic'' mesh spacing parameter $h$: The static mesh discussed in Sec.~\ref{sec:staticResults} may not be optimal to represent  the collective wave function. As time evolves, the collective wave function is for instance subject to strong variations in the first potential well. Describing accurately these fluctuations often requires a finer mesh than the one used to describe the smooth variations of the potential. To address this issue, we perform the time evolution on a different grid from the static PES. The typical procedure used to build this auxiliary grid is described in~\cite{regnier2016}. It starts from a regular grid entirely defined by the parameter $h$ through: $\Delta_{q_{20}}=h$ fm$^2$, $\Delta_{q_{30}}=3h$ fm$^3$. After a Delaunay triangulation this initial mesh is h-refined in two areas, namely the inner and outer potential wells and their saddle points as well as a band lying along the frontier line. Subsequent p-refinement over the whole domain provides a locally refined mesh composed of quadratic Lagrange elements.
\item The time step $\delta t$: This parameter characterizes the time discretization of the evolution equation (\ref{eq:tdgcmgoa}).
\item A tolerance for the matrix inversions $\epsilon_\text{inv}$: At each time step, a linear system is solved iteratively. This parameter corresponds to the maximal residual tolerated by the inversion method.
\item A parameter $\epsilon_\text{init}$ controlling the numerical precision of the initial state: The determination of the initial state is based on the search for quasi-bound states, which requires diagonalizing the auxiliary Hamiltonian $\hat{H}'$ of Eq.(\ref{eq:eigenStates}). The parameter $\epsilon_{\text{init}}$ characterizes the numerical precision when determining the eigenvectors of $\hat{H}'$.
\end{itemize}

\begin{table}[!ht]
\caption{Convergence of fission yields in the whole mass range $A_H\in[120,160]$ as a function of the time step $\delta t$ (left part) and the grid size parameter $h$ (right part). For each parameter, the error is defined as $\epsilon= \text{Sup}\{(Y(A)-Y_r(A))/Y_r(A) | A\in[120,160]\}$, where $Y_r(A)$ corresponds to the most refined calculation (e.g. $h=300$ in the case of the mesh space parameter). Time is expressed in zepto-seconds (1 zs = $10^{-21}$ s). Values marked in boldface correspond to the choice (\ref{eq:numericalParam}).}
\begin{ruledtabular}
\begin{tabular}{ccc|cc}
$\delta t$ (zs)  & $\epsilon$ (\%) & & $h$ (fm units) & $\epsilon$ (\%)    \\
4.10$^{-3}$      & 9.1        & &      848       &     38        \\ 
2.10$^{-3}$      & 6.7        & &      600       &     10        \\ 
\textbf{1.10$^{-3}$} & \bf{1.9} &  & \bf{424}  &   \bf{2}   \\
5.10$^{-4}$      & 0          & &      300       &     0       \\
\end{tabular}
\end{ruledtabular}
\label{tab:convergence}
\end{table}

All the results presented in this paper are obtained with the following set of parameters, unless specified otherwise:
\begin{equation}
\label{eq:numericalParam}
\begin{array}{l}
 h= 424 \text{ (fm units)},\ \delta t= 10^{-3} \text{ (zs, 1 zs = $10^{-21}$ s)}, \\
 \epsilon_{\text{inv}}=10^{-15},\ \epsilon_{\text{init}}=10^{-13}
\end{array}
\end{equation}
The tolerance $\epsilon_{\text{inv}}$ and $\epsilon_{\text{init}}$  are chosen so that the errors related to the linear system inversions and the diagonalization are several orders of magnitude below any other source of numerical error. The cumulative effect of inversion errors can be quantified by the evolution of the norm of the collective wave function in calculations without any absorption condition. For our choice of $\epsilon_{\text{inv}}$, we obtain a relative error on the norm smaller than $10^{-12}$ after 16 zs. As for $\epsilon_{\text{init}}$, we verified that going from $1.10^{-12}$ to $1.10^{-14}$ impacts the fission yields by a maximum of $2.10^{-6}$ \% only.
The convergence with respect to the space and time parameters was checked by performing series of calculations in the vicinity of the set (\ref{eq:numericalParam}). For each parameter, we explored a wide range of values while the other one remained unchanged. As an example, we report in table~\ref{tab:convergence} the convergence of the D1S fission yields. From this results, we can expect a numerical relative precision of a few percents in the whole mass region $A\in[120,160]$. Additional discussion on the numerical precision of our TDGCM+GOA FELIX solver can be found in \cite{regnier2016}.


\begin{figure}[!ht]
\includegraphics[width=0.45\textwidth]{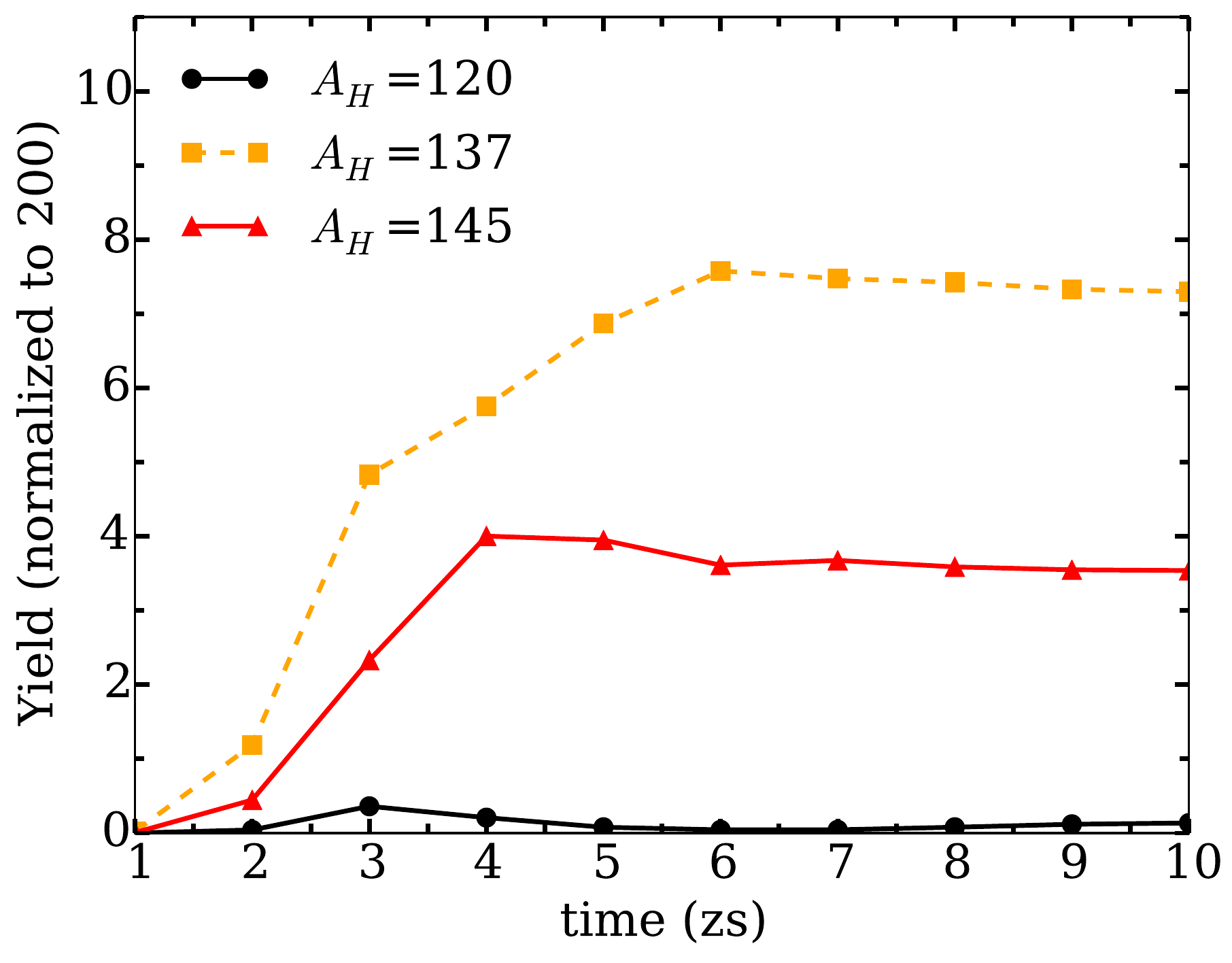}
\caption{(color online) Evolution of the SkM* heavy mass yields as a function of time for three possible fragmentations: $A_H=120$ (symmetric valley), $A_H=137$ (asymmetric peak), $A_H=145$ (very asymmetric wing of the distribution).}
\label{fig:yA_tConvergence.pdf}
\end{figure}

\subsection{Fission yields obtained with SkM* and D1S}

In this section, we present the pre-neutron fragment yields for \Pu239(n,f) computed with the SkM* and D1S EDF. The initial state consists of  a Gaussian superposition of quasi-bound states with average energy at 1 MeV above the inner barrier.
The solver computes the wave packet evolution up to a time $t_{\text{max}}$ determined by the relation
\begin{equation}
 \frac{1}{F_{\text{tot}}(t_{\text{max}})} \left. \frac{\partial F_{\text{tot}}}{\partial t}\right | _{t_{\text{max}}} < 10^{-3} \text{ zs}^{-1},
\end{equation}
where $F_{\text{tot}}(t)$ is the total flux crossing the frontier. With this criterion, the solver computes the dynamics up to 10 zs with SkM* and 16 zs with D1S. Such a difference may typically arise from the change in potential and inertia of the two EDF around the outer saddle point. Figure \ref{fig:yA_tConvergence.pdf} shows that after a transient regime of roughly 7 zs, the SkM* fission yields stabilize. The variations of the yields in the last zepto-second of simulation are of the order of 9.2\% in the symmetric valley ($A_H=120$) and 0.3\% in the asymmetric peak ($A_H\simeq137$) as well as in the very asymmetric wings of the distribution ($A_H\simeq145$). The D1S time evolution shows a similar behavior with a transient regime of 10 zs followed by a plateau of the yields. 

Within this lapse of time, 10\% (SkM*) and 18\% (D1S) of the total wave packet norm crosses the frontier. As expected, the  wave packet follows essentially the asymmetric fission valley. In the case of D1S, the endpoint of this valley, located in the region $Q_{30} \in [25, 50]$ b$^{3/2}$ (\textit{i.e} $\xi \in [85,112]$), accounts for 90\% of the outgoing flux. The SkM* case gives similar results with 91\% of the outgoing flux in the same $Q_{30}$ range (which corresponds this time to $\xi \in [65,92]$).

\begin{figure}[!ht]
\includegraphics[width=0.45\textwidth]{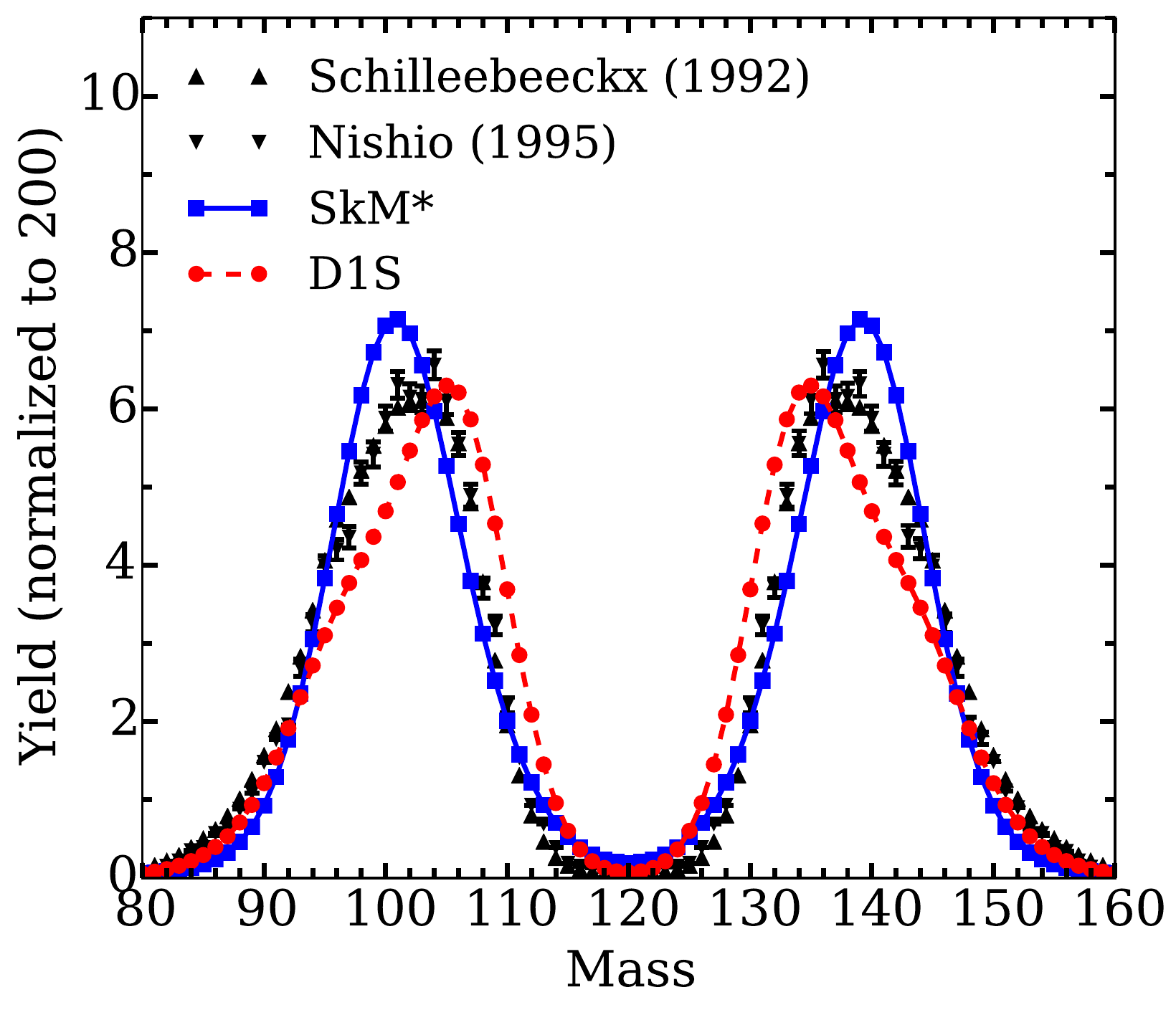}
\caption{(color online) Pre-neutron mass yields for \Pu239(n,f). The SkM* and D1S calculations are compared with two experimental datasets~\cite{schillebeeckx_comparative_1992,nishio_measurement_1995}. The data from Nishio~\etal \ are plotted with their statistical uncertainties.}
\label{fig:yA_interaction}
\end{figure}

In figure~\ref{fig:yA_interaction}, we compare the mass yields predictions versus two experimental data sets. 
The asymmetric nature of the \Pu240 fission is well described by the TDGCM approach with both EDF. We show the main characteristics of the asymmetric peaks in table~\ref{tab:yieldsProp}. The centroids of the peaks agree within 2 mass units between theory and experiment. The SkM* EDF favors slightly more asymmetric configurations whereas we observe the opposite for D1S. Note that the shift of the D1S peak toward symmetry is consistent with earlier work by Younes and Gogny relying on a different collective space and a totally different numerical implementation both of the DFT and TDGCM+GOA solver \cite{younes2012-a}; see also \cite{goutte2005} for details of the TDGCM+GOA solver used then. The widths of the predicted mass yields match the experiment within roughly 1 mass unit. Part of this agreement comes from the fact that both calculations and experimental data are convolutions of a raw data set with a smoothing function: The Schillebeeckx \etal~\cite{schillebeeckx_comparative_1992} data comes from a 2E measurement with a mass resolution of typically 4-5 mass units (FWHM)~\cite{wagemans_nuclear_1991}. A measurement of both energy and velocity of the two fragments as the one performed by Nishio \etal~\cite{nishio_measurement_1995} is also expected to provide a mass resolution significantly larger than one mass unit. These experimental characteristics partly motivated our choice of a $\sigma=4$ Gaussian smoothing function, together with the arguments presented in Sec.~\ref{sec:fragmentDistribution}.

\begin{table}[!ht]
\caption{Characteristics of the fission pre-neutron mass yields for \Pu239(n,f). The notation $A_{H,p}$ accounts for the position of the heavy peak and the full width at half height is denoted $\Sigma_{A}$.}
\begin{ruledtabular}
\begin{tabular}{lcccc}
     & $A_{H,p}$ & $Y(A_{H,p})$ (\%) & $\Sigma_{A}$ & $Y(A_{H,p})/Y(120)$ \\
SkM*           & 139 & 7.1 & 25 & 36.8 \\
D1S            & 135 & 6.3 & 24 & 87.4 \\
Schillebeeckx  & 137 & 6.1 & 26 & 105.9 \\
Nishio         & 136 & 6.6 & 25 & 72.9 \\
\end{tabular}
\end{ruledtabular}
\label{tab:yieldsProp}
\end{table}

Finally, our results show strong variations of the peak-to-valley ratio as a function of the EDF. However, a detailed comparison of this peak-to-valley ratio seems premature, since at least three features may impact our results for the symmetric valley. First of all, this mass region has been shown to be highly sensitive to the parity of the initial state~\cite{goutte2005}. In our calculations, we have made no effort to match such parity with information on the neutron entrance channel. Secondly, the presence of persistent discontinuities in the symmetric part of the collective landscape may impact the yields around mass $A_H =120$. Finally, the low probability of symmetric fragmentations induces large experimental uncertainties of the corresponding fission yields.

\begin{figure}[!ht]
\includegraphics[width=0.45\textwidth]{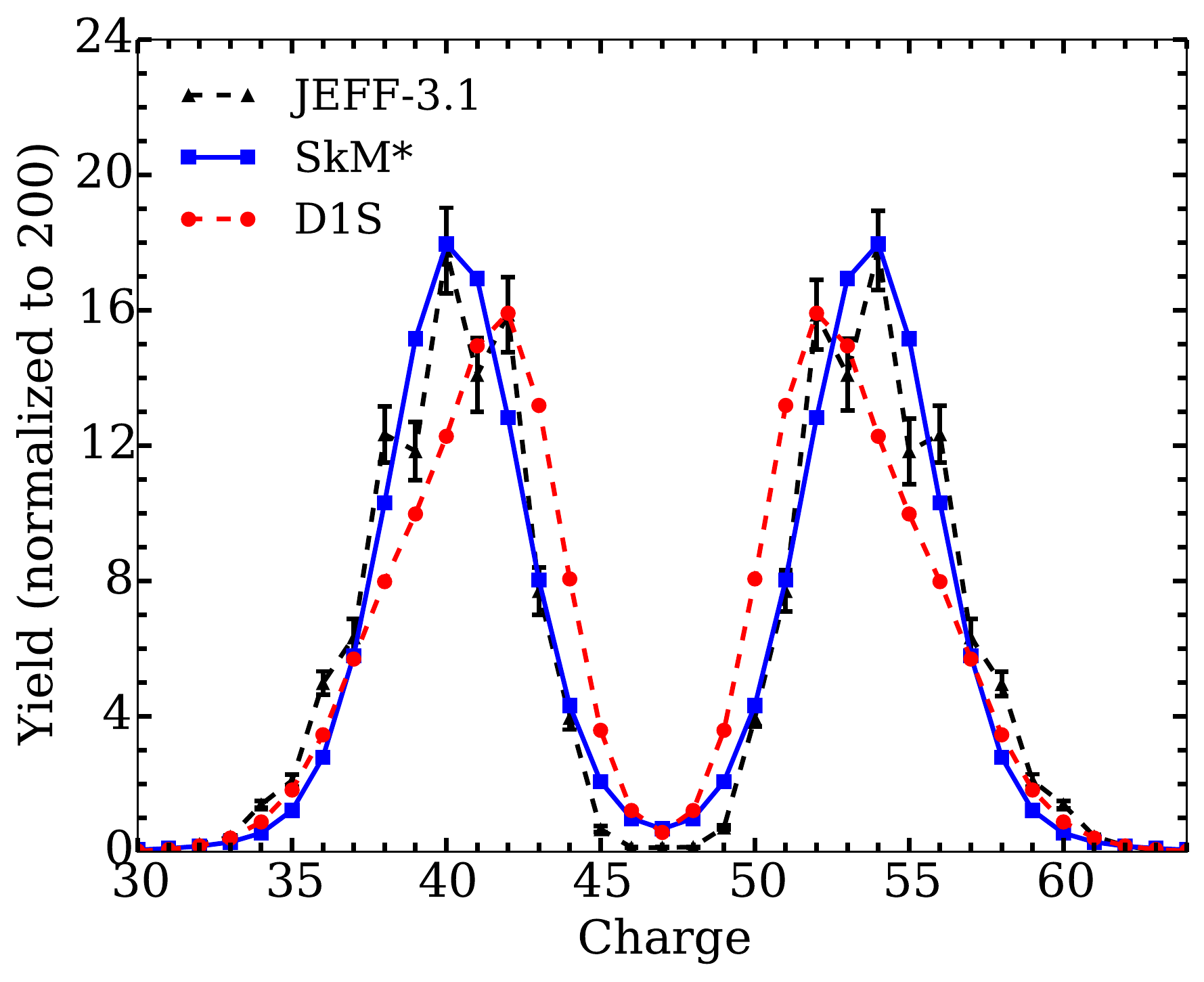}
\caption{(color online) Charge yields for \Pu239(n,f). The calculation obtained with a SkM* functional (plain line) is compared to the European evaluation JEFF-3.1 (dashed line).}
\label{fig:yZ_skmVersusExp}
\end{figure}

Figure~\ref{fig:yZ_skmVersusExp} compares the charge yields obtained in the same calculations with the results from the European JEFF-3.1 evaluation. Overall, we observe the same trends as for the mass yields: The SkM* force favors more asymmetric configurations whereas we observe the opposite for D1S. Note that our model cannot account for odd-even staggering effects. This would require either projection of particle number for the fragment and/or the use of collective variables corresponding to different charge channels in analogy to what is done in macroscopic-microscopic methods \cite{moeller2015}. Without the use of such fragment-specific constraint, the potential energy surface does not contain enough information to describe odd-even features of fragment distributions.


\subsection{\label{subsec:sensitivity}  Sensitivity Analysis}

While the agreement between theory and experiment reported in figure \ref{fig:yA_interaction} seems very promising, one should be clear that our model depends on several input ``parameters'' that may have an influence on our predictions. In this section, we provide a first estimate of the related uncertainties. 


\subsubsection{\label{subsubsec:resInitialState} Sensitivity to the initial state}

\begin{figure}[!ht]
\includegraphics[width=0.45\textwidth]{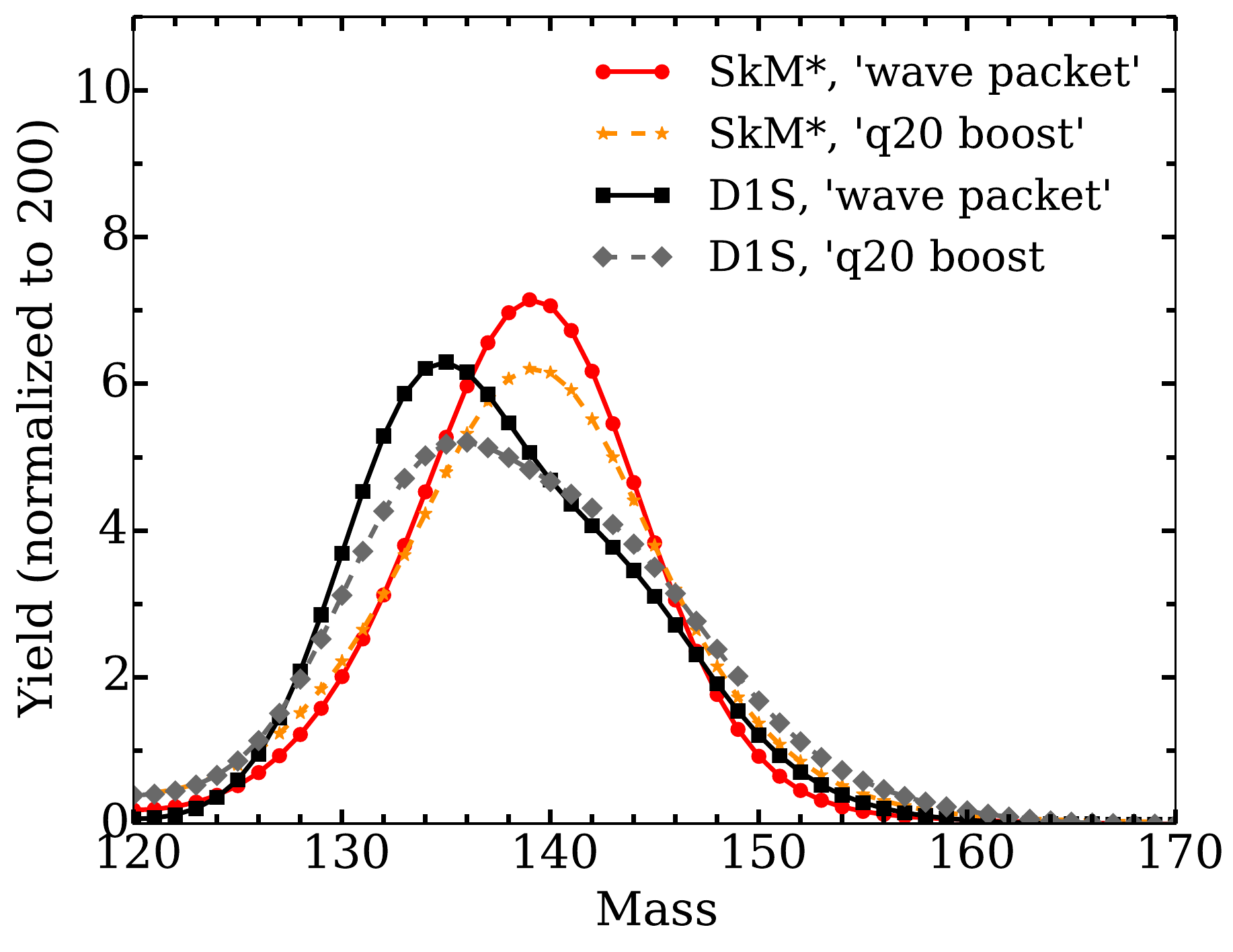}
\caption{(color online) Pre-neutron mass yields for \Pu239(n,f) obtained from the two different initial states described in Sec.~\ref{subsubsec:initialState}. Calculations were performed with both SkM* and D1S potential energy surfaces. Only the heavy mass region is represented.}
\label{fig:yA_initState}
\end{figure}

We performed TDGCM+GOA calculations starting from the two different initial states described in~\ref{subsubsec:initialState}. The figure~\ref{fig:yA_initState} shows the fission yields obtained with both D1S and SkM* interactions for these different scenarios. The '$\hat{Q}_{20}$ boosted' state concentrates most of the kinetic energy of the system in the degree of freedom associated with the elongation. It results in an increase of the peak to valley ratio  by a factor of 6.7 compared to the 'wave packet' initial state. Typically, the change in initial state also induces a 20\% variation of the yields at the peak whereas no pronounced mass shift is observed.

Our results indicate that a phenomenological model of the initial state of the compound nucleus can quantitatively reproduce the main features of the fission product distributions as long as the average energy and average deformation of the initial wave packet are realistic. At this point, one may choose between two options. In the short term, one may decide to retain such a phenomenological description of the initial state, and use various statistical methods to determine the ``best'' guess for the initial state and quantify the related uncertainties. This could be achieved by comparing the mass distributions with experimental data for a few actinide nuclei. In the short term, this simple approach would be very useful for applications where a precision higher than 20\% on the overall distributions is needed, as it would enable uncertainty propagation methods. In the long term, however, one should try to avoid such phenomenology (which defeats the purpose of developing a microscopic approach) and model directly the initial state by using a proper theory of the entrance channel.


\subsubsection{Comparison with the ATDHFB formalism}

As recalled in section \ref{subsubsec:collectiveFields}, the GCM approach to computing the collective inertia tensor is known to be deficient unless collective variables associated with the collective momentum are explicitly taken into account. As a result, fission studies often rely on the ATDHFB prescription for the collective inertia \cite{staszczak2013,rodriguez-guzman2014,rodriguez-guzman2014-a,sadhukhan2014,baran2015}. Let us briefly recall that this semi-classical formalism is extensively presented in~\cite{brink1976,villars1977,baranger1978}. It leads to a classical Hamiltonian in terms of the collective coordinates and related moments. Using the Pauli quantization~\cite{krappe2012} procedure yields a local Schr\"odinger-like equation with a collective Hamiltonian which is formally identical to Eq.~(\ref{eq:Hcoll}). The differences with the GCM approach are in the definition of the components $V$, $\Bvec$, $\gamma$ of the Hamiltonian. As discussed extensively in the literature (see for instance~\cite{schunck2016} and references therein), the construction of the full ATDHFB inertia tensor involves time-odd states of the system and the values are about 1.5 larger than the GCM inertia. In addition, the Pauli quantization does not produce zero point corrections in the potential part, so that the collective potential is simply given by
\begin{equation}
V(\qvec)= \langle \Phi_\qvec | \hat{H} | \Phi_\qvec \rangle.
\end{equation}
Finally, the metric appearing in the collective Schr\"odinger equation is simply the determinant of ATHDF inertia mass tensor instead of being related to the metric tensor $\Gvec$, which is itself a function of the overlap between generator states,
\begin{equation}
\gamma(\qvec)=\operatorname{det}\left(M_{\text{ATDHFB}}(\qvec)\right).
\end{equation}

To gain some insight on the dependence of the fission fragment yields on the collective inertia, we solved the quantized ATDHFB equations for both the D1S and SkM* effective interactions. The initial states are again defined as wave packets of quasi-bound states. Due to the difference in inertia, the ATDHFB spectrum of the quasi-bound states is compressed compared to GCM. As a consequence, the energy window $[B_{I}; B_{I}+2\text{ MeV}]$, which is crucial to define our initial collective wave packet, contains a different set of states. For example, with the D1S interaction the ATDHFB initial wave packet is mainly built out of the 15 states numbered from 20 to 28 in the spectrum, while in the GCM case only 6 states from the 11$^{\text{th}}$ to the 15$^{\text{th}}$ are available in the same energy window.

\begin{figure}[!ht]
\includegraphics[width=0.45\textwidth]{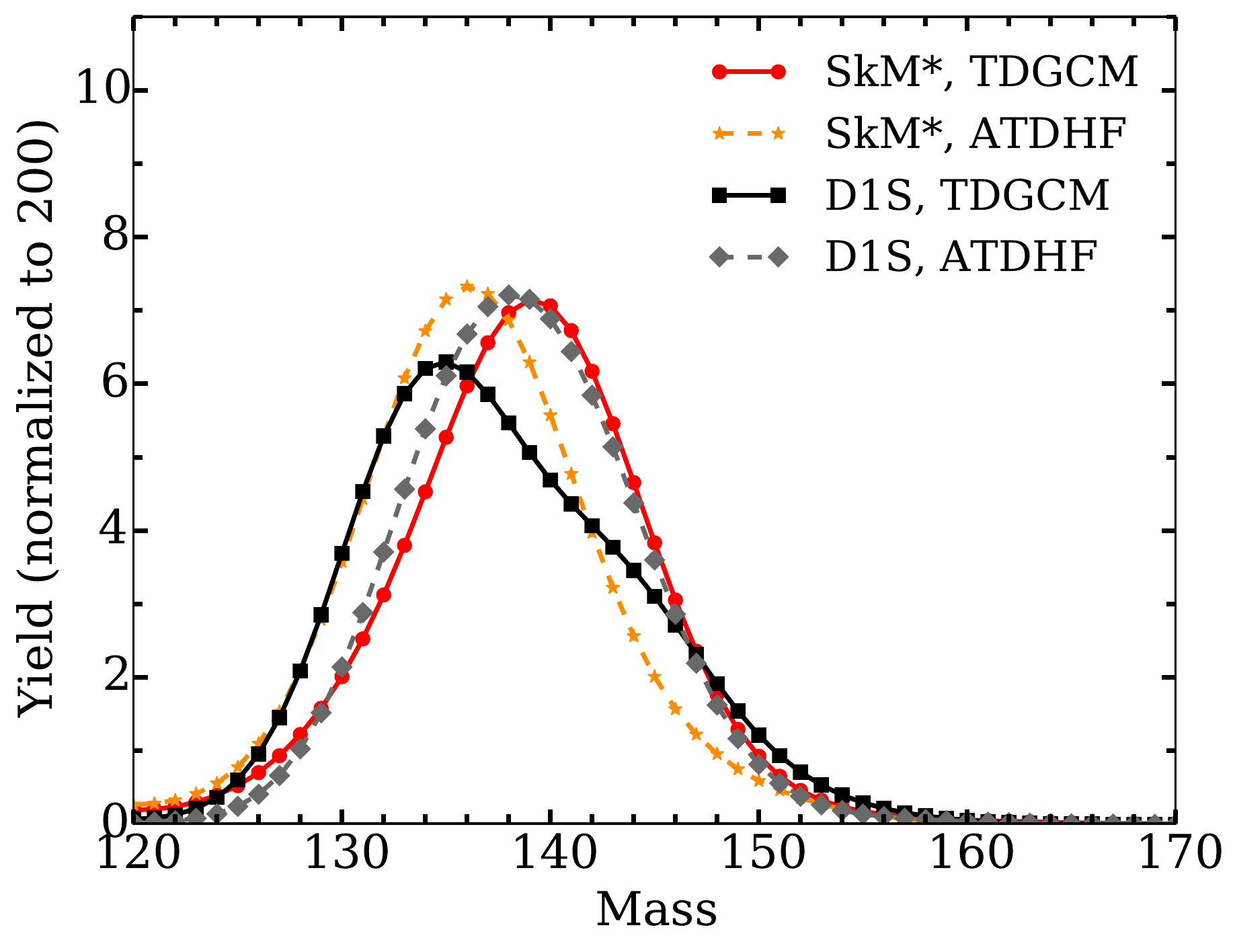}
\caption{(color online) Pre-neutron mass yields for \Pu239(n,f) obtained in the strict TDGCM and quantized ATDHFB formalisms. Calculations were performed with both SkM* and D1S potential energy surfaces using the same prescription for the initial state. Only the heavy mass region is represented.}
\label{fig:yA_inertia}
\end{figure}

Figure~\ref{fig:yA_inertia} shows the mass yields of the heavy fragments for the two prescriptions for the inertia tensor. Both SkM* and D1S results are sensitive to this change in inertia. However, the deviation between the TDGCM and ATDHFB calculations has a different pattern for the two EDF. In the D1S case, going from the TDGCM to the ATDHFB formalism induces a 3 mass units shift of the peak toward asymmetry and an increase of the peak yield by 15 \%. In the case of SkM*, the change in inertia causes a 3 mass unit shift of the peak toward symmetry. These opposite behaviors reflect the fact that the collective inertia is involved in several aspects of the time-evolution. As discussed above, it actually impacts the definition of the initial state for the dynamics. In addition, we also expect that changes of inertia will favor different areas of the collective space as the system evolves in time. Finally, the inertia enters explicitly the expression of the collective flux (cf. Eq.~\ref{eq:fluxDef}).


\subsubsection{\label{subsec:gaussian} Sensitivity to the convolution width}

\begin{figure}[!ht]
\includegraphics[width=0.45\textwidth]{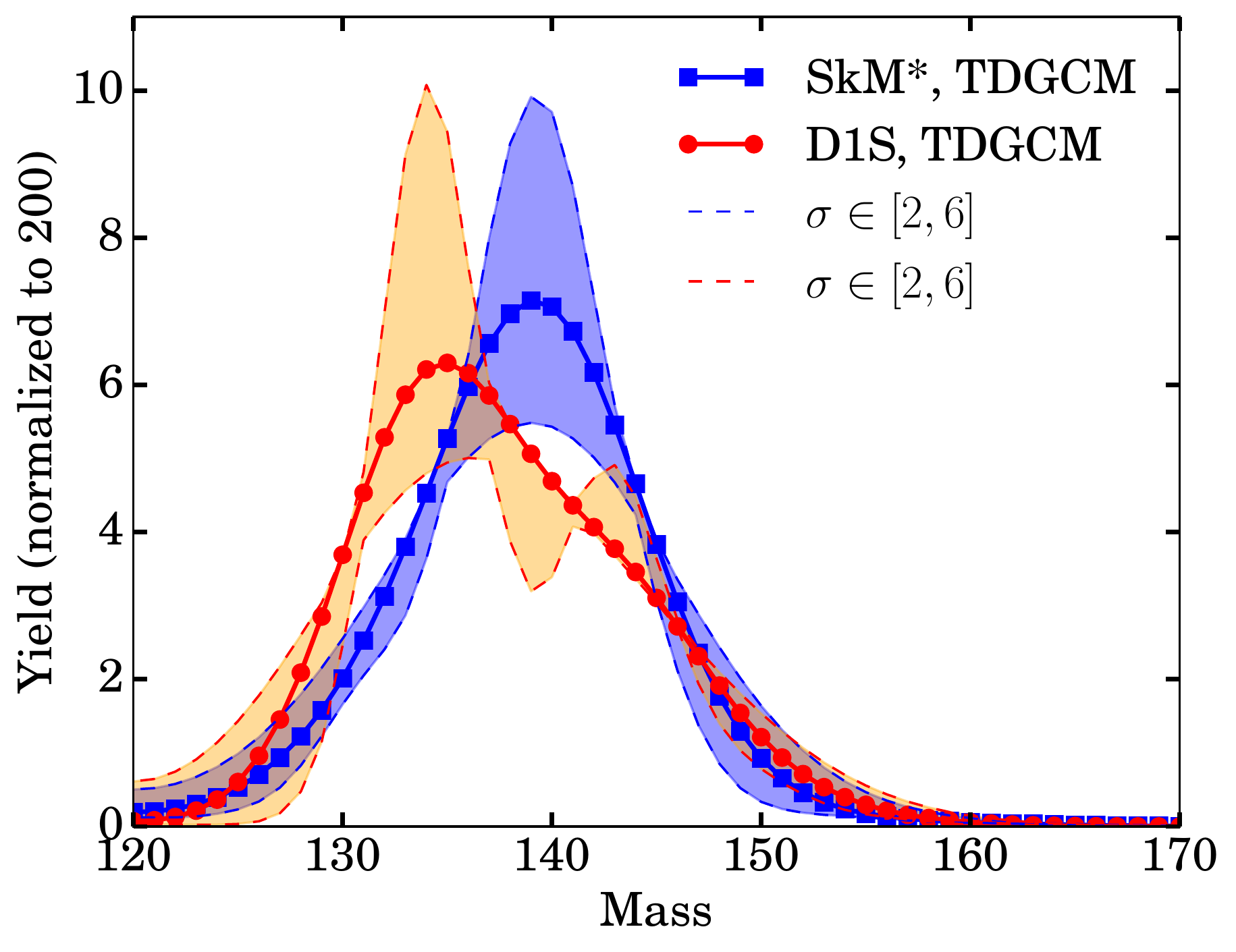}
\caption{(color online) Sensitivity of the pre-neutron mass yields for \Pu239(n,f) to the width of the Gaussian convolution. Calculations are based on the strict TDGCM formalism.}
\label{fig:yA_sigma}
\end{figure}

As discussed in Sec.~\ref{sec:fragmentDistribution}, the calculated fission yields result from the convolution of the raw flux through the frontier with a Gaussian. Although this choice is motivated by several physics considerations, such a convolution remains phenomenological and our choice for the value of the Gaussian width ($\sigma=4$ amu) may be discussed. To estimate the influence of this parameter, we performed a series of convolutions with different widths ranging from 2 to 6 mass units. The figure~\ref{fig:yA_sigma} summarizes the sensitivity of the mass yields to such a variation. We observe strong variations of the yields reaching a factor of 1.7 at the peak. On the other hand, the full width at half height remains mainly constant in this range of $\sigma$. For D1S, the small structure of the uncertainty band near $A_H=138$ is dominated by the convolution with the smallest width $\sigma=2$. This is the trademark of the fast variation of the fragmentations in the area $\xi \in [95,105]$ of the frontier (see Fig.~\ref{fig:frontierQn4}). These error bands give us an indication of the gains in precision that would come from a formalism able to describe the dynamics up to two well-separated output channels.

\subsubsection{\label{subsec:triaxiality} Impact of triaxiality}

It is well-known that triaxiality plays a major role in the low deformation region of the PES and especially on the inner barrier height \cite{girod1983,bender1998,warda2002,staszczak2005,staszczak2009,abusara2010,lu2012,schunck2014}. To estimate its effect on the dynamics, we compare the wave packet propagation for the SkM* EDF based on two PES computed with and without breaking axial symmetry. In practice, a constraint on $\hat{Q}_{22}$ was imposed, and subsequently released, in the region $Q_{20}<300$ b and $Q_{30}<76$ b$^{3/2}$, which allows the HFODD DFT solver to explore possible triaxial configurations.

Based on this new PES, we repeated the construction of the initial state as a wave packet of quasi-bound states with an energy 1 MeV above the first barrier. By construction, the excitation energy of the corresponding compound nucleus decreases from 10.3 MeV (axial) down to 8.7 MeV when triaxiality is included. The FELIX solver then propagates the collective wave function up to 12 zs (compared to 10 zs for the axial case). This small increase in scission time is not surprising since the system possesses a smaller energy whereas the outer barrier has remained unchanged. At the end of the simulation, we find that the fission yields obtained are nearly identical to the axial case. The relative differences are only 11 \% over the whole region of heavy mass fragments, $A_H\in[127,150]$, with small variations of the yields in the symmetric valley at a maximum of 17 \% for $A=121$. These results suggest that, while triaxiality has a strong impact on the absolute probability to fission, it does not seem to fundamentally change the relative population of the output channels. 


\section{\label{sec:conclusion} Conclusion}

We have presented a benchmark calculation of the mass and charge distributions for low energy \Pu239(n,f) within the framework of the TDGCM+GOA. We detailed the numerical and technical issues arising in both the static and dynamic parts of this type of calculation. Our analysis confirms that the adiabatic approximation provides an effective scheme to compute fission fragment yields. The main characteristics of the fission charge and mass distributions can be well reproduced by existing energy functionals even in the two-dimensional collective space spanned by the quadrupole and octupole moments. In addition, we highlighted the sensitivity of our predictions to variations of its different inputs. Overall, we found that the qualitative features of fission fragment mass distribution are rather robust and independent of the EDF and/or the various ingredients of the model. This should facilitate the validation of the predictive power of the method on different fissioning systems, especially in actinide nuclei where quality data exists. 

On the other hand, this study also highlights several shortcomings of our implementation of the DFT+TDGCM approach. In particular, the nagging issue of discontinuities in the PES suggests that more collective variables may need to be included in the calculation, especially near scission. Note that discontinuities may also be avoided by changing the definition of the collective variables following the work by Younes and Gogny \cite{younes2012-a}. In the long term, it seems also clear that a better modeling of several important physics effects, such as the description of the initial state, of collective inertia, of particle number both in the fissioning system and in the fragments, may be necessary to reach the level of 10\% accuracy in predicting pre-neutron yields needed by applications in science and technology. A more comprehensive description of the fission process capable among others of quantitatively predicting the evolution of the yields as a function of the neutron incident energy may also require going beyond the hypothesis of adiabaticity assumed in the present paper.


\begin{acknowledgments}
Part of this research was performed under the auspices of the U.S.\ Department of Energy by Lawrence Livermore National Laboratory under Contract DE-AC52-07NA27344. Computational resources were provided through an INCITE award ``Computational Nuclear Structure" by the National Center for Computational Sciences (NCCS). Computing resources were also provided through an award by the Livermore Computing Resource Center at Lawrence Livermore National Laboratory.
\end{acknowledgments}


\bibliographystyle{apsrev4-1}
\bibliography{zotero_output,books,unpublished,biblio}

\end{document}